\pdfoutput=1

\documentclass[12pt,a4paper]{article}

\usepackage{ifthen} 
\newboolean{pdflatex}
\setboolean{pdflatex}{true} 

\newboolean{articletitles}
\setboolean{articletitles}{true} 

\newboolean{uprightparticles}
\setboolean{uprightparticles}{false} 

\newboolean{inbibliography}
\setboolean{inbibliography}{false} 

\usepackage[top=1in, bottom=1.25in, left=1in, right=1in]{geometry}

\usepackage{microtype}
\usepackage{lineno}  
\usepackage{xspace} 
\usepackage{caption} 

\usepackage{graphicx}  
\usepackage{color}
\usepackage{colortbl}
\graphicspath{{./figs/}} 

\usepackage{amsmath} 
\usepackage{amssymb}
\usepackage{amsfonts}
\usepackage{upgreek} 

\newcommand*\patchAmsMathEnvironmentForLineno[1]{%
\expandafter\let\csname old#1\expandafter\endcsname\csname #1\endcsname
\expandafter\let\csname oldend#1\expandafter\endcsname\csname
end#1\endcsname
 \renewenvironment{#1}%
   {\linenomath\csname old#1\endcsname}%
   {\csname oldend#1\endcsname\endlinenomath}%
}
\newcommand*\patchBothAmsMathEnvironmentsForLineno[1]{%
  \patchAmsMathEnvironmentForLineno{#1}%
  \patchAmsMathEnvironmentForLineno{#1*}%
}
\AtBeginDocument{%
\patchBothAmsMathEnvironmentsForLineno{equation}%
\patchBothAmsMathEnvironmentsForLineno{align}%
\patchBothAmsMathEnvironmentsForLineno{flalign}%
\patchBothAmsMathEnvironmentsForLineno{alignat}%
\patchBothAmsMathEnvironmentsForLineno{gather}%
\patchBothAmsMathEnvironmentsForLineno{multline}%
\patchBothAmsMathEnvironmentsForLineno{eqnarray}%
}

\usepackage{hyperref}    
\usepackage[all]{hypcap} 

\usepackage{xspace} 
\usepackage{upgreek}

\def\lhcb {\mbox{LHCb}\xspace}

\def\MagUp {\mbox{\em Mag\kern -0.05em Up}\xspace}

\ifthenelse{\boolean{uprightparticles}}%
{

 \def\Pmu         {\ensuremath{\upmu}\xspace}                 
 \def\Pnu         {\ensuremath{\upnu}\xspace}                 
                  
 \def\Ppi         {\ensuremath{\uppi}\xspace}

 \def\Ppsi        {\ensuremath{\uppsi}\xspace}

 \def\PDelta      {\ensuremath{\Delta}\xspace}                 
 \def\PXi      {\ensuremath{\Xi}\xspace}                 
 \def\PLambda      {\ensuremath{\Lambda}\xspace}                 
 \def\PSigma      {\ensuremath{\Sigma}\xspace}                 
 \def\POmega      {\ensuremath{\Omega}\xspace}                 
 \def\PUpsilon      {\ensuremath{\Upsilon}\xspace}                 
 
 \def\PB      {\ensuremath{\mathrm{B}}\xspace}                 
                  
 \def\PD      {\ensuremath{\mathrm{D}}\xspace}

 \def\PJ      {\ensuremath{\mathrm{J}}\xspace}                 
 \def\PK      {\ensuremath{\mathrm{K}}\xspace}

 \def\Pb      {\ensuremath{\mathrm{b}}\xspace}                 
 \def\Pc      {\ensuremath{\mathrm{c}}\xspace}

 \def\Pi      {\ensuremath{\mathrm{i}}\xspace}

 \def\Pp      {\ensuremath{\mathrm{p}}\xspace}

 \def\Ps      {\ensuremath{\mathrm{s}}\xspace}

}
{

 \def\Pmu         {\ensuremath{\mu}\xspace}                 
 \def\Pnu         {\ensuremath{\nu}\xspace}                 
                  
 \def\Ppi         {\ensuremath{\pi}\xspace}

 \def\Ppsi        {\ensuremath{\psi}\xspace}                 
                  
 \mathchardef\PDelta="7101
 \mathchardef\PXi="7104
 \mathchardef\PLambda="7103
 \mathchardef\PSigma="7106
 \mathchardef\POmega="710A
 \mathchardef\PUpsilon="7107
                  
 \def\PB      {\ensuremath{B}\xspace}                 
                  
 \def\PD      {\ensuremath{D}\xspace}

 \def\PJ      {\ensuremath{J}\xspace}                 
 \def\PK      {\ensuremath{K}\xspace}

 \def\Pb      {\ensuremath{b}\xspace}                 
 \def\Pc      {\ensuremath{c}\xspace}

 \def\Pi      {\ensuremath{i}\xspace}

 \def\Pp      {\ensuremath{p}\xspace}

 \def\Ps      {\ensuremath{s}\xspace}

}

\makeatletter
\ifcase \@ptsize \relax
  \newcommand{\miniscule}{\@setfontsize\miniscule{4}{5}}
\or
  \newcommand{\miniscule}{\@setfontsize\miniscule{5}{6}}
\or
  \newcommand{\miniscule}{\@setfontsize\miniscule{5}{6}}
\fi
\makeatother

\DeclareRobustCommand{\optbar}[1]{\shortstack{{\miniscule (\rule[.5ex]{1.25em}{.18mm})}
  \\ [-.7ex] $#1$}}

\def\mup        {{\ensuremath{\Pmu^+}}\xspace}
\def\mun        {{\ensuremath{\Pmu^-}}\xspace} 

\def\neu        {{\ensuremath{\Pnu}}\xspace}

\def\neum       {{\ensuremath{\neu_\mu}}\xspace}

\def\squark    {{\ensuremath{\Ps}}\xspace}

\def\cquark    {{\ensuremath{\Pc}}\xspace}

\def\bquark    {{\ensuremath{\Pb}}\xspace}
\def\bquarkbar {{\ensuremath{\overline \bquark}}\xspace}

\def\pion   {{\ensuremath{\Ppi}}\xspace}
\def\piz    {{\ensuremath{\pion^0}}\xspace}

\def\pip    {{\ensuremath{\pion^+}}\xspace}
\def\pim    {{\ensuremath{\pion^-}}\xspace}

\def\kaon    {{\ensuremath{\PK}}\xspace}
  \def\Kbar    {{\kern 0.2em\overline{\kern -0.2em \PK}{}}\xspace}

\def\KorKbar    {\kern 0.18em\optbar{\kern -0.18em K}{}\xspace}

\def\Kp      {{\ensuremath{\kaon^+}}\xspace}
\def\Km      {{\ensuremath{\kaon^-}}\xspace}

\def\KS      {{\ensuremath{\kaon^0_{\mathrm{ \scriptscriptstyle S}}}}\xspace}

\def\Kstarz  {{\ensuremath{\kaon^{*0}}}\xspace}
\def\Kstarzb {{\ensuremath{\Kbar{}^{*0}}}\xspace}
\def\Kstar   {{\ensuremath{\kaon^*}}\xspace}

  \def\Dbar    {{\kern 0.2em\overline{\kern -0.2em \PD}{}}\xspace}
\def\D       {{\ensuremath{\PD}}\xspace}

\def\DorDbar    {\kern 0.18em\optbar{\kern -0.18em D}{}\xspace}
\def\Dz      {{\ensuremath{\D^0}}\xspace}
\def\Dzb     {{\ensuremath{\Dbar{}^0}}\xspace}
\def\Dp      {{\ensuremath{\D^+}}\xspace}

\def\Dstarb  {{\ensuremath{\Dbar{}^*}}\xspace}
\def\Dstarz  {{\ensuremath{\D^{*0}}}\xspace}

\def\Dstarp  {{\ensuremath{\D^{*+}}}\xspace}

\def\Ds      {{\ensuremath{\D^+_\squark}}\xspace}
\def\Dsp     {{\ensuremath{\D^+_\squark}}\xspace}

\def\Dssp    {{\ensuremath{\D^{*+}_\squark}}\xspace}

\def\B       {{\ensuremath{\PB}}\xspace}
\def\Bbar    {{\ensuremath{\kern 0.18em\overline{\kern -0.18em \PB}{}}}\xspace}

\def\BorBbar    {\kern 0.18em\optbar{\kern -0.18em B}{}\xspace}

\def\Bu      {{\ensuremath{\B^+}}\xspace}

\def\Bp      {{\ensuremath{\Bu}}\xspace}

\def\Bs      {{\ensuremath{\B^0_\squark}}\xspace}

\def\Bc      {{\ensuremath{\B_\cquark^+}}\xspace}
\def\Bcp     {{\ensuremath{\B_\cquark^+}}\xspace}

\def\jpsi     {{\ensuremath{{\PJ\mskip -3mu/\mskip -2mu\Ppsi\mskip 2mu}}}\xspace}
\def\psitwos  {{\ensuremath{\Ppsi{(2S)}}}\xspace}

  \def\Y#1S{\ensuremath{\PUpsilon{(#1S)}}\xspace}

\def\proton      {{\ensuremath{\Pp}}\xspace}
\def\antiproton  {{\ensuremath{\overline \proton}}\xspace}

\def\Lbar        {{\ensuremath{\kern 0.1em\overline{\kern -0.1em\PLambda}}}\xspace}
\def\LorLbar    {\kern 0.18em\optbar{\kern -0.18em \PLambda}{}\xspace}

\def\BF         {{\ensuremath{\mathcal{B}}}\xspace}

\def\to                 {\ensuremath{\rightarrow}\xspace}

\def\AT#1     {\ensuremath{A_{\mathrm{T}}^{#1}}\xspace}           

\def\C#1      {\ensuremath{\mathcal{C}_{#1}}\xspace}                       
\def\Cp#1     {\ensuremath{\mathcal{C}_{#1}^{'}}\xspace}                    
\def\Ceff#1   {\ensuremath{\mathcal{C}_{#1}^{\mathrm{(eff)}}}\xspace}        
\def\Cpeff#1  {\ensuremath{\mathcal{C}_{#1}^{'\mathrm{(eff)}}}\xspace}       
\def\Ope#1    {\ensuremath{\mathcal{O}_{#1}}\xspace}                       
\def\Opep#1   {\ensuremath{\mathcal{O}_{#1}^{'}}\xspace}                    

\newcommand{\tev}{\ifthenelse{\boolean{inbibliography}}{\ensuremath{~T\kern -0.05em eV}\xspace}{\ensuremath{\mathrm{\,Te\kern -0.1em V}}}\xspace}
\newcommand{\gev}{\ensuremath{\mathrm{\,Ge\kern -0.1em V}}\xspace}
\newcommand{\mev}{\ensuremath{\mathrm{\,Me\kern -0.1em V}}\xspace}
\newcommand{\kev}{\ensuremath{\mathrm{\,ke\kern -0.1em V}}\xspace}
\newcommand{\ev}{\ensuremath{\mathrm{\,e\kern -0.1em V}}\xspace}
\newcommand{\gevc}{\ensuremath{{\mathrm{\,Ge\kern -0.1em V\!/}c}}\xspace}
\newcommand{\mevc}{\ensuremath{{\mathrm{\,Me\kern -0.1em V\!/}c}}\xspace}
\newcommand{\gevcc}{\ensuremath{{\mathrm{\,Ge\kern -0.1em V\!/}c^2}}\xspace}
\newcommand{\gevgevcccc}{\ensuremath{{\mathrm{\,Ge\kern -0.1em V^2\!/}c^4}}\xspace}
\newcommand{\mevcc}{\ensuremath{{\mathrm{\,Me\kern -0.1em V\!/}c^2}}\xspace}

\def\mum  {\ensuremath{{\,\upmu\mathrm{m}}}\xspace}

\def\invfb   {\ensuremath{\mbox{\,fb}^{-1}}\xspace}

\def\fs   {\ensuremath{\mathrm{ \,fs}}\xspace}

\newcommand{\stat}{\ensuremath{\mathrm{\,(stat)}}\xspace}
\newcommand{\syst}{\ensuremath{\mathrm{\,(syst)}}\xspace}

\newcommand{\chisq}{\ensuremath{\chi^2}\xspace}
\newcommand{\chisqndf}{\ensuremath{\chi^2/\mathrm{ndf}}\xspace}
\newcommand{\chisqip}{\ensuremath{\chi^2_{\text{IP}}}\xspace}

\def\gsim{{~\raise.15em\hbox{$>$}\kern-.85em
          \lower.35em\hbox{$\sim$}~}\xspace}
\def\lsim{{~\raise.15em\hbox{$<$}\kern-.85em
          \lower.35em\hbox{$\sim$}~}\xspace}

\def\ptot       {\mbox{$p$}\xspace}
\def\pt         {\mbox{$p_{\mathrm{ T}}$}\xspace}

\def\evtgen     {\mbox{\textsc{EvtGen}}\xspace}

\def\geant      {\mbox{\textsc{Geant4}}\xspace}

\def\photos     {\mbox{\textsc{Photos}}\xspace}

\def\pythia     {\mbox{\textsc{Pythia}}\xspace}

\def\tell1  {TELL1\xspace}
\def\ukl1   {UKL1\xspace}

\newcommand{\Dmesons}{\D^{(*)}}
\newcommand{\Kmesons}{\kaon^{(*)}}
\newcommand{\Dneutrals}{\D^{(*)0}}

\usepackage{cite} 
\usepackage{mciteplus}

\usepackage{longtable} 

\begin{document}

\renewcommand{\thefootnote}{\fnsymbol{footnote}}
\setcounter{footnote}{1}


\begin{titlepage}
\pagenumbering{roman}

\vspace*{-1.5cm}
\centerline{\large EUROPEAN ORGANIZATION FOR NUCLEAR RESEARCH (CERN)}
\vspace*{1.5cm}
\noindent
\begin{tabular*}{\linewidth}{lc@{\extracolsep{\fill}}r@{\extracolsep{0pt}}}
 & & CERN-EP-2016-303 \\  
 & & LHCb-PAPER-2016-055 \\  
 & & 20 March 2017 \\ 
 & & \\
\end{tabular*}

\vspace*{4.0cm}

{\normalfont\bfseries\boldmath\huge
\begin{center}
  Observation of $B_c^+ \to J/\psi D^{(*)} K^{(*)}$ decays
\end{center}
}

\vspace*{2.0cm}

\begin{center}
The LHCb collaboration\footnote{Authors are listed at the end of this paper.}
\end{center}

\vspace{\fill}

\begin{abstract}
  \noindent
  A search for the decays $\Bcp \to \jpsi D^{(*)0} \Kp$ and 
  $\Bcp \to \jpsi D^{(*)+} \Kstarz$ is performed with data collected at the LHCb
  experiment 
  corresponding to an integrated luminosity of 3\invfb.
  The decays $\Bcp \to \jpsi \Dz \Kp$ and $\Bcp\to\jpsi \Dstarz K^+$ 
  are observed for the first time, while first evidence is reported for
  the $\Bcp \to \jpsi D^{*+} \Kstarz$ and $\Bcp \to \jpsi \Dp \Kstarz$ decays.
  The branching fractions of these decays are determined 
  relative to the $\Bcp \to \jpsi \pip$ decay.
  The \Bc mass is measured, using the $\jpsi \Dz \Kp$ final state, to be 
  $6274.28 \pm 1.40 \stat \pm 0.32 \syst\mevcc$. This is the most precise 
  single measurement of the $\Bcp$ mass to date.

\end{abstract}

\vspace*{2.0cm}

\begin{center}
  Published in Phys.~Rev.~D 95, 032005 (2017)
\end{center}

\vspace{\fill}

{\footnotesize 
\centerline{\copyright~CERN on behalf of the \lhcb collaboration, licence \href{http://creativecommons.org/licenses/by/4.0/}{CC-BY-4.0}.}}
\vspace*{2mm}

\end{titlepage}


\newpage
\setcounter{page}{2}
\mbox{~}

\cleardoublepage

\renewcommand{\thefootnote}{\arabic{footnote}}
\setcounter{footnote}{0}



\pagestyle{plain} 
\setcounter{page}{1}
\pagenumbering{arabic}


\section{Introduction}
\label{sec:intro}
\begin{figure}[htb]
\includegraphics[width=0.33\textwidth]{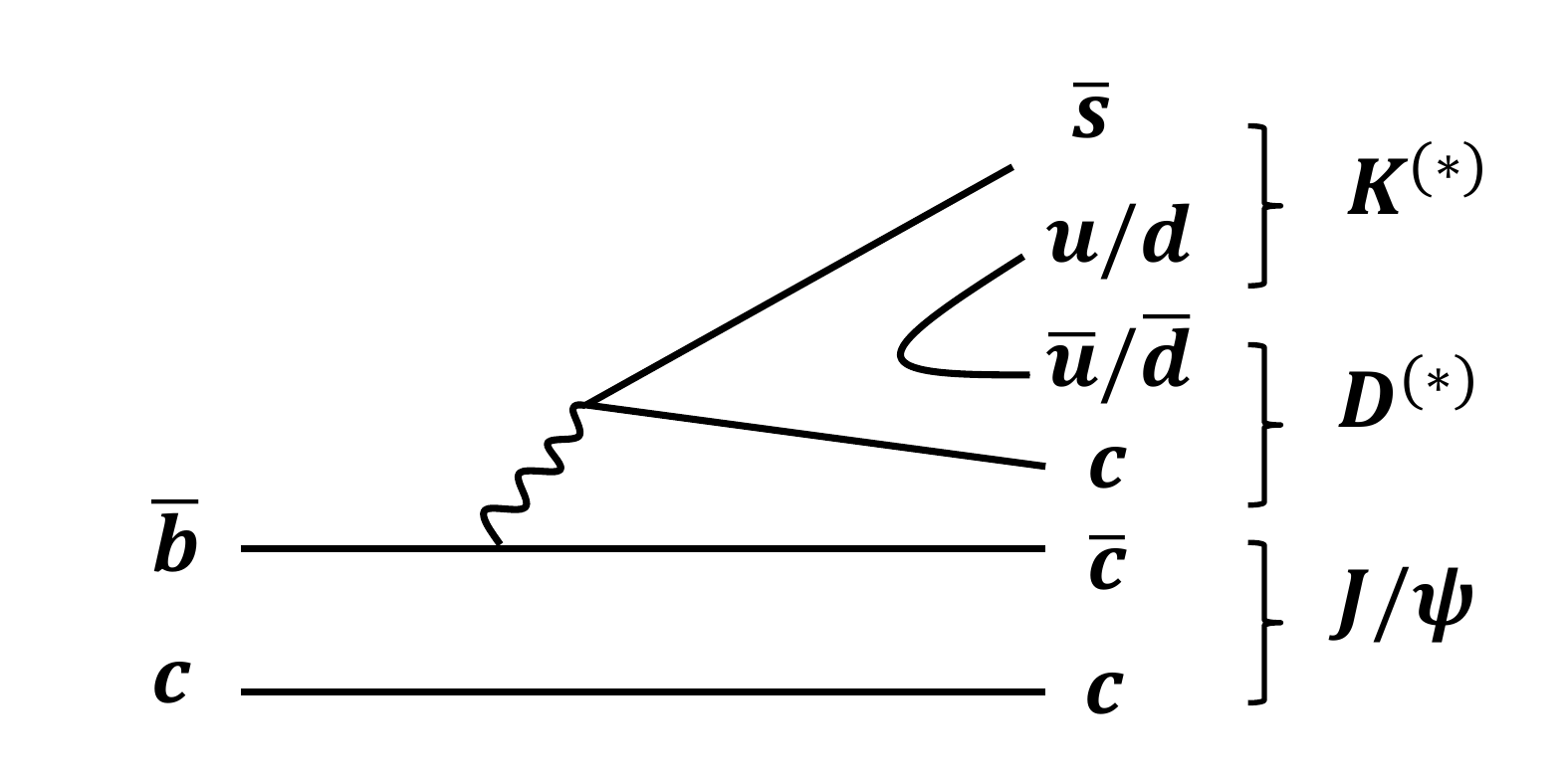}
\includegraphics[width=0.32\textwidth]{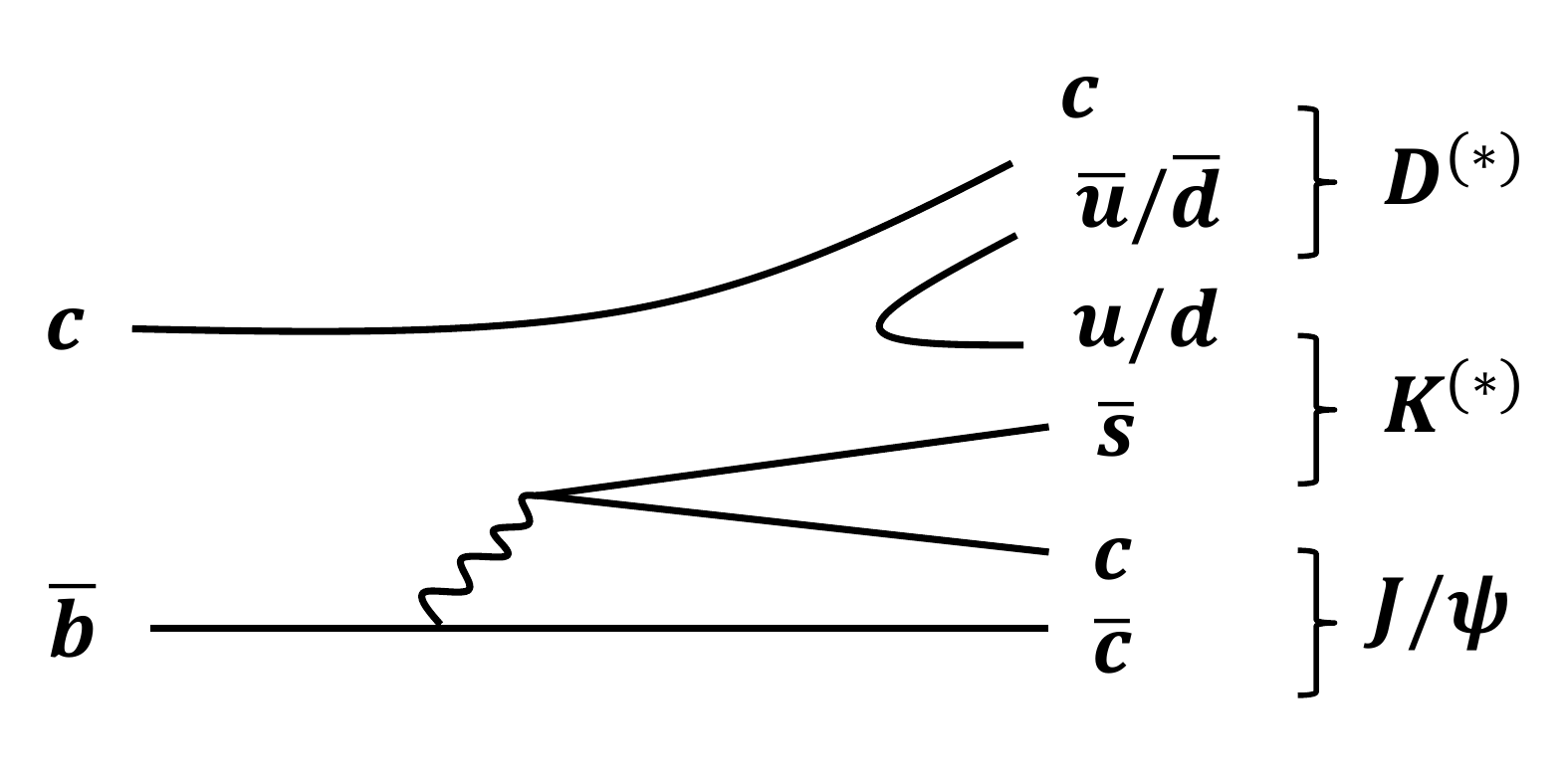}
\includegraphics[width=0.31\textwidth]{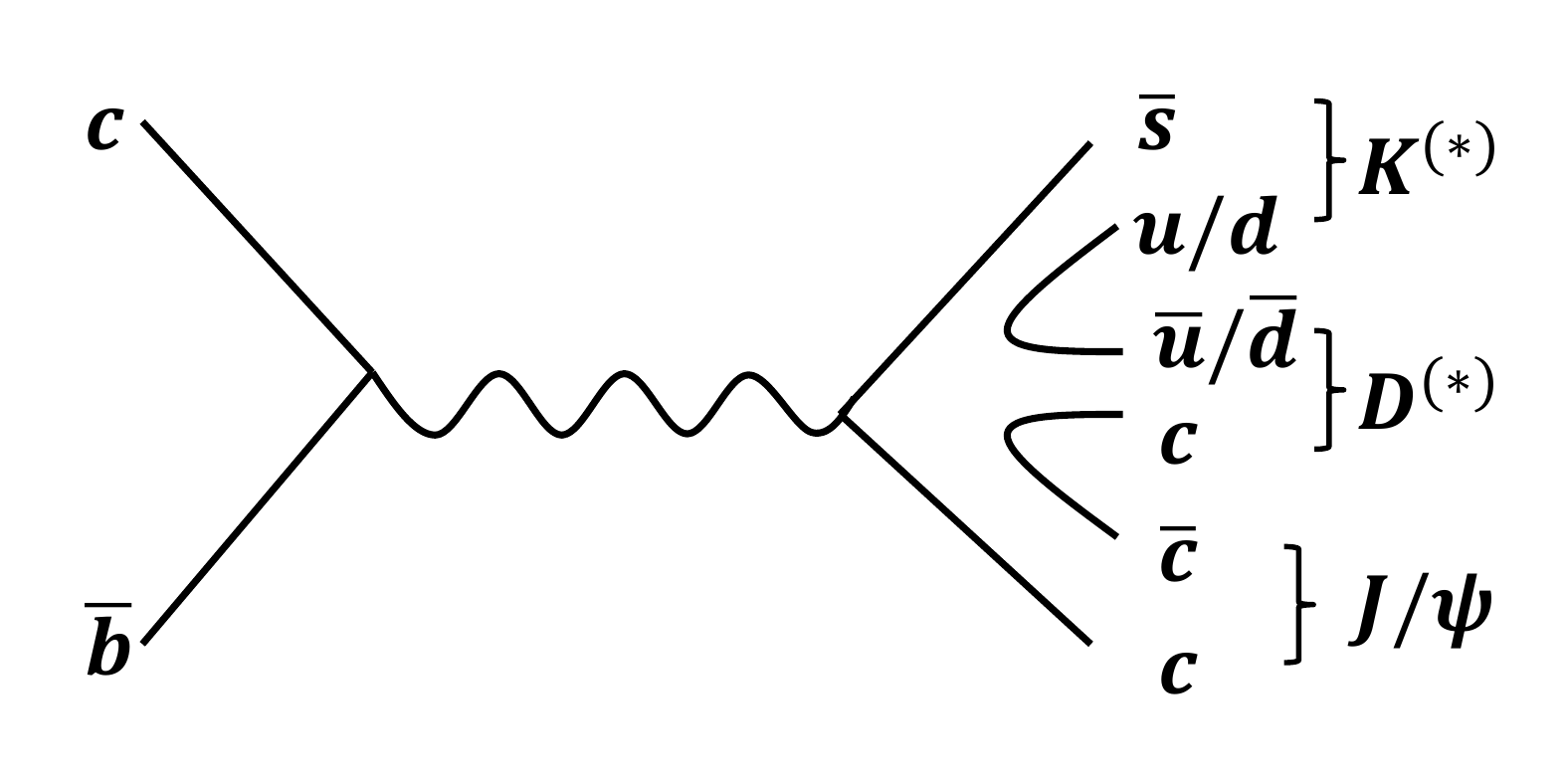}
\caption{Diagrams for $\Bcp\to\jpsi\Dmesons\Kmesons$ decays mediated by $\bquarkbar \to \cquark \bar{c} \bar{s}$ and annihilation amplitudes. }
\label{fig:feynman}
\end{figure}

Composed of two heavy quarks of different flavour, the \Bcp meson is 
the least understood member of the pseudoscalar bottom-meson family.
The high centre-of-mass energies 
at the Large Hadron Collider enable the LHCb experiment to study 
the production, properties and decays of the $\Bcp$ meson\footnote{The inclusion of charge-conjugate processes is implied throughout.}~\cite{LHCb-PAPER-2012-028,LHCb-PAPER-2014-050,LHCb-PAPER-2013-044,LHCb-PAPER-2013-063,LHCb-PAPER-2014-060,LHCb-PAPER-2011-044,LHCb-PAPER-2012-054,LHCb-PAPER-2015-024,LHCb-PAPER-2013-021,LHCb-PAPER-2013-010,LHCb-PAPER-2013-047,LHCb-PAPER-2014-009,LHCb-PAPER-2014-039,LHCb-PAPER-2014-025}.
As for the $\Bcp \to \jpsi D_\squark^{(*)+}$ decays~\cite{LHCb-PAPER-2013-010}, the
  $\Bcp \to \jpsi\Dmesons\Kmesons$ decays are expected to proceed mainly through spectator diagrams. In contrast to decays of other beauty hadrons, the weak annihilation
topology is not suppressed and can contribute significantly to the decay amplitude (Fig.~\ref{fig:feynman}).
The $\Bcp~\to~\jpsi\Dmesons\Kmesons$ decays offer a unique opportunity to
study \Ds spectroscopy in the $\Dmesons\Kmesons$ 
system~\cite{Fu:2011tn,Sun:2015uva}. Given a large enough sample size,
the quantum numbers of possible excited $\D_{\squark J}^+$ states can be determined, 
complementary to inclusive 
searches~\cite{LHCb-PAPER-2012-016,LHCb-PAPER-2015-052}
and Dalitz analyses of other $B$ meson 
decays~\cite{LHCb-PAPER-2014-035,LHCb-PAPER-2014-036}.
The complex structure of the $\Bcp \to \jpsi \Dmesons \Kmesons$ decay also 
allows the search for exotic charmonium states 
in the $\jpsi\Dmesons$ combination.
A measurement of the relative branching fraction 
$\BF(\Bcp\to\jpsi\Dmesons \Kstar)/\BF(\Bcp\to\jpsi\Dmesons \kaon)$ provides information
on the branching fraction of the as yet unobserved $B \to \Dstarb\Dmesons \Kstar$ decay, in which
exotic charmonia close to the $\Dstarb\Dmesons$ threshold can be studied.
The search for $\Bcp\to\jpsi\Dmesons\Kmesons$ decays in this paper is a first
step towards such spectroscopy studies.

The current world average of the $\Bcp$ mass measurements~\cite{PDG2016} is
dominated by the LHCb results using $\jpsi\pip$~\cite{LHCb-PAPER-2012-028},
$\jpsi\Dsp$~\cite{LHCb-PAPER-2013-010} and 
$\jpsi \proton \antiproton \pip$~\cite{LHCb-PAPER-2014-039} decays. 
The $\jpsi\pip$ measurement benefits from a large yield while the latter 
two have smaller systematic uncertainties because of their reduced 
$Q$-values.\footnote{The $Q$-value is defined as the difference between 
the mass of the parent particle and the sum of the masses of its decay products.}
With a $Q$-value even smaller than the $\Bcp \to \jpsi\Dsp$ or $\jpsi \proton \antiproton \pip$ channels,
the $\Bcp \to \jpsi\Dz\Kp$ decay enables another precise 
$\Bcp$ mass measurement.

The purpose of this analysis is to search for the $\Bcp$ meson decaying into the final
states $\jpsi \Dz \Kp$, $\jpsi \Dstarz \Kp$, $\jpsi \Dp \Kstarz$
and $\jpsi \Dstarp \Kstarz$.
The $\Dz$ meson is reconstructed in both $\Km\pip$ and $\Km\pip\pim\pip$ 
final states in the search for the $\Bcp \to \jpsi\Dneutrals\Kp$ decays,
and only in the $\Km\pip$ final state for the other decays.
The $\Dp$ meson is reconstructed in the $\Km\pip\pip$ final state.
The decays $\Dstarz \to \Dz\gamma$, $\Dstarz \to \Dz\piz$, and $\Dstarp \to \Dz\pip$ are partially reconstructed retaining only the \Dz while neglecting the photon or pion.
The $\jpsi$ is reconstructed in the $\mup\mun$ final state.
The relative branching fraction of the $\Bcp\to\jpsi\Dz(\to\Km\pip)\Kp$ decay
is measured with respect to the $\Bcp \to \jpsi \pip$ decay, 
while the other channels are normalised to the $\Bcp \to \jpsi \Dz \Kp$ decay.
The determination of the $\Bcp$ mass is performed with the 
$\Bcp\to \jpsi\Dz(\to \Km\pip)\Kp$ final state only.

\section{Detector and dataset}
\label{sec:detector}
This analysis uses $pp$ collision data collected at the LHCb experiment corresponding to 
an integrated luminosity of $1.0 \invfb$ at a centre-of-mass energy 
of $7\tev$ and $2.0 \invfb$ at $8\tev$.
The \lhcb detector~\cite{Alves:2008zz,LHCb-DP-2014-002} is a single-arm forward
spectrometer covering the \mbox{pseudorapidity} range $2<\eta <5$,
designed for the study of particles containing \bquark or \cquark
quarks. The detector includes a high-precision tracking system
consisting of a silicon-strip vertex detector surrounding the $pp$
interaction region, a large-area silicon-strip detector located
upstream of a dipole magnet with a bending power of about
$4{\mathrm{\,Tm}}$, and three stations of silicon-strip detectors and straw
drift tubes placed downstream of the magnet.
The polarity of the dipole magnet is reversed periodically 
throughout data taking.
The tracking system provides a measurement of momentum, \ptot, 
of charged particles with a relative uncertainty that varies 
from 0.5\% at low momentum to 1.0\% at $200\gevc$.
The minimum distance of a track to a primary vertex (PV), the impact parameter 
(IP), is measured with a resolution of $(15+29/\pt)\mum$,
where \pt is the component of the momentum transverse to the beam, in\gevc.
Different types of charged hadrons are distinguished using information
from two ring-imaging Cherenkov detectors.
Photons, electrons and hadrons are identified by a calorimeter system 
consisting of scintillating-pad and preshower detectors, an electromagnetic
calorimeter and a hadronic calorimeter. Muons are identified by a
system composed of alternating layers of iron and multiwire
proportional chambers.

The online event selection is performed by a trigger,
which consists of a hardware stage, based on information from the calorimeter 
and muon systems, followed by a software stage, which applies a full event
reconstruction.
For all decays considered in this paper, a trigger is used that enriches
events with $\jpsi$ decays into the two-muon final state.
At the hardware trigger level the signal candidates are required to 
contain at least one muon with $\pt > 1.48\gevc$ ($> 1.76\gevc$)
in the 7\tev (8\tev) data, or a muon pair where the product of the $\pt$ 
values of the muons is greater than $(1.3\gevc)^2$ and $(1.6\gevc)^2$ 
in the 7\tev and 8\tev data, respectively.
In the first step of the software trigger a single 
muon candidate with $\pt > 1.0\gevc$ is required, or a pair
of oppositely charged muons, each with $\pt > 500\mevc$, 
with a combined invariant mass $M_{\mu\mu}> 2.7\gevcc$.
Finally, a $\jpsi$ candidate is required to be formed 
from a muon pair, and to have a mass within $\pm120\mevcc$ of the known $\jpsi$
mass~\cite{PDG2016} and a vertex position displaced from its associated PV
with a significance of at least three standard deviations ($\sigma$).

Simulated samples of the signal and the normalisation channel
are used to optimise the selection criteria and to estimate 
the efficiencies. The simulation of $\Bcp$ production in $pp$ collisions is modelled 
with the \textsc{BcVegPy} generator~\cite{Chang:2003cq,Chang:2005hq},
interfaced to \mbox{\pythia 6}~\cite{Sjostrand:2006za}
with a specific \lhcb configuration~\cite{LHCb-PROC-2010-056}. 
Decays of hadronic particles are described by \evtgen~\cite{Lange:2001uf}, 
in which final-state radiation is generated using \photos~\cite{Golonka:2005pn}.
The interaction of the generated particles with the detector, and its response,
are implemented using the \geant 
toolkit~\cite{Allison:2006ve, *Agostinelli:2002hh} as described in
Ref.~\cite{LHCb-PROC-2011-006}.

\section{Event selection}
\label{sec:selection}
The offline selection starts with a loose preselection and is followed
by a multivariate selection using a boosted decision tree 
(BDT)~\cite{Breiman,Roe}. This is done independently for each of 
the final states considered: 
\begin{itemize}
\item{$\jpsi\Dz(\to \Km\pip)\Kp$};
\item{$\jpsi\Dz(\to \Km\pip\pim\pip)\Kp$};
\item{$\jpsi\Dz(\to \Km\pip) \Kstarz(\to \Kp\pim)$};
\item{$\jpsi\Dp(\to \Km\pip\pip) \Kstarz(\to \Kp\pim)$};
\item{$\jpsi\pip$ (normalisation channel).}
\end{itemize}
In the offline selection, trigger decisions are associated with reconstructed 
particles.
In order to establish whether a significant signal is observed
no requirements are placed on whether
the trigger decision is due to the signal candidate itself or 
other particles in the event.
In the branching fraction and mass measurements 
it is required that the trigger decision must be 
due to the signal candidate (denoted TOS, Trigger-On-Signal) for a better determination 
of the trigger efficiency.

In the preselection each $\jpsi$ candidate is formed from a pair of muons, 
each with a good-quality track fit, $\pt$ in excess of $550\mevc$, and 
minimum $\chisqip$ with respect to any reconstructed PV greater than 4, 
where $\chisqip$ is the difference between the vertex-fit $\chisq$ of 
a given PV reconstructed with and without the considered track.
The $\chisqip$ requirement rejects tracks that come from the associated PV rather than 
from $\Bcp$ decays, where the associated PV is the primary
vertex\footnote{The majority of the data has in average 1.8 visible interactions per beam-beam crossing.} with respect to which the $\Bcp$ candidate has the smallest $\chisqip$.
The muons are required to be positively identified with neural-network-based 
particle identification (PID) variables using information 
from different sub-detectors.
The muon pair is required to form a vertex of good quality and 
have an invariant mass in the range $3040$--$3150\mevcc$.
The $\jpsi$ candidate is then combined with hadron tracks
to form a $\Bcp$ candidate.
All hadronic tracks are required to have a good-quality track fit, 
$\pt$ in excess of $100\mevc$, and the minimum $\chisqip$ with respect to any PV greater than 4. 
Loose PID requirements are applied to pions and kaons for the
$\jpsi\Dz(\to\Km\pip)\Kp$ final state, while tighter selections on kaons
are applied at a later stage.
For other final states, tighter PID selections are imposed in the preselections.
The \Dz and \Dp candidates are required to have a good-quality vertex, and
have a mass within $\pm 30\mevcc$ of the known masses,
where the size of the window corresponds to approximately
 $\pm 4$ times the mass resolution.
The $\Kstarz$ meson is defined as a $\Kp\pim$ combination within the mass range 
$792$--$992\mevcc$, roughly four times the $K^{*}(892)^0$ natural 
width~\cite{PDG2016}. 
The $\Bcp$ candidate is required to have a good-quality vertex and 
a mass within a wide window $\pm 700\mevcc$ around the world average $\Bcp$ mass~\cite{PDG2016}.

A BDT discriminator is trained for each of the signal final states 
to further suppress the combinatorial background, 
except that the partially reconstructed $\jpsi\Dstarz\Kp$ decay
shares the same BDT as the fully reconstructed $\jpsi\Dz\Kp$ decay.
The training uses simulated samples as signal, 
and background events from data containing $\Kmesons$ candidates of opposite strangeness as in the 
respective signal decays
(for example, $\jpsi\Dz\Km$ for $\jpsi\Dz\Kp$
signal, or $\jpsi\Dp\Kstarzb$ for $\jpsi\Dp\Kstarz$ signal, later referred to 
as ``wrong-sign" samples).
Taking the $\jpsi\Dz(\to\Km\pip)\Kp$ decay as an example,
the variables used in the training fall into the following categories:
\begin{itemize}
\item{the $\pt$ of the $\Bcp$ candidate and its decay products: 
$\jpsi$, $\Dz$ and $\Kp$;}
\item{vertex-fit $\chisq$ per degree of freedom ($\chisqndf$) of the $\Bcp$, 
$\jpsi$ and $\Dz$ mesons, as well as 
$\chisqndf$ from a refit of the $\Bcp$ decay constraining 
 the reconstructed $\jpsi$ and $\Dz$ masses to their known values, and 
 the $\Bcp$ momentum to point back to its associated PV;}
\item{variables describing the event geometry: 
the flight distance significances (FDS) of the $\Bcp$ and $\Dz$ candidates 
with respect to its associated PV, 
where FDS is the distance between the vertex and the reference point
divided by its uncertainty;
$\chisqip$ and $\theta$ of the $\Bcp$ meson relative to its associated PV, 
where $\theta$ is the angle between the $\Bcp$ momentum and the line connecting
its production vertex and decay vertex;
$\chisqip$ and $\theta$ of the $\Dz$ meson relative to the $\Bcp$  decay vertex;
$\Dz$ decay length from the refit with constraints mentioned above.}
\end{itemize}
For other final states, the variables corresponding to the $\Dz$ or $\Kp$ mesons are
replaced with those corresponding to the $\Dp$ or $\Kstarz$ mesons as appropriate.

The thresholds of the BDT discriminants are chosen to maximise the 
figure of merit $\varepsilon/(3/2 + \sqrt{N_B})$~\cite{Punzi:2003bu},
aiming for a signal significance of three standard deviations,
where $\varepsilon$ is the signal efficiency estimated from simulation
and $N_B$ is the number of expected background candidates in the signal region
($6263$--$6289\mevcc$ for fully reconstructed signals, and $6037$--$6149\mevcc$ 
for the partially reconstructed $\Bcp\to\jpsi\Dstarp\Kstarz$ decay),
extrapolated from the wrong-sign samples.
For the $\jpsi\Dz(\to\Km\pip)\Kp$ final state 
the BDT discriminant output and the PID variables of the kaons
are optimised simultaneously,
while for the other final states only the BDT discriminant is optimised 
since tighter PID selections have already been imposed.
When there is more than one candidate present in a selected event, 
the one with the smallest $\chisqndf$ in the constrained vertex refit 
is retained.

For the normalisation channel $\Bcp\to\jpsi\pip$, the training variables 
are similar to the signal channels, except for the absence of variables 
related to the $\Dz$ meson, 
and the addition of the pion $\pt$ and $\chisqip$.
Simulated signal decays are used in the training, while the background
sample is taken from signal candidates in the upper sideband
($6500 \leq M(\jpsi\pip) \leq 6800\mevcc$) in data.
The BDT discriminant is chosen to maximise the signal significance
$N_S/\sqrt{N_S + N_B}$, where $N_S$ is the expected signal yield, and 
$N_S + N_B$ is the total number of candidates in the region
$6241$--$6312\mevcc$ corresponding to $\pm 3$ times the mass resolution 
around the $\Bcp$ mass.

\section{Signal yields}
\label{sec:yields}
\begin{figure}[t]
\centering
\includegraphics[width=0.48\textwidth]{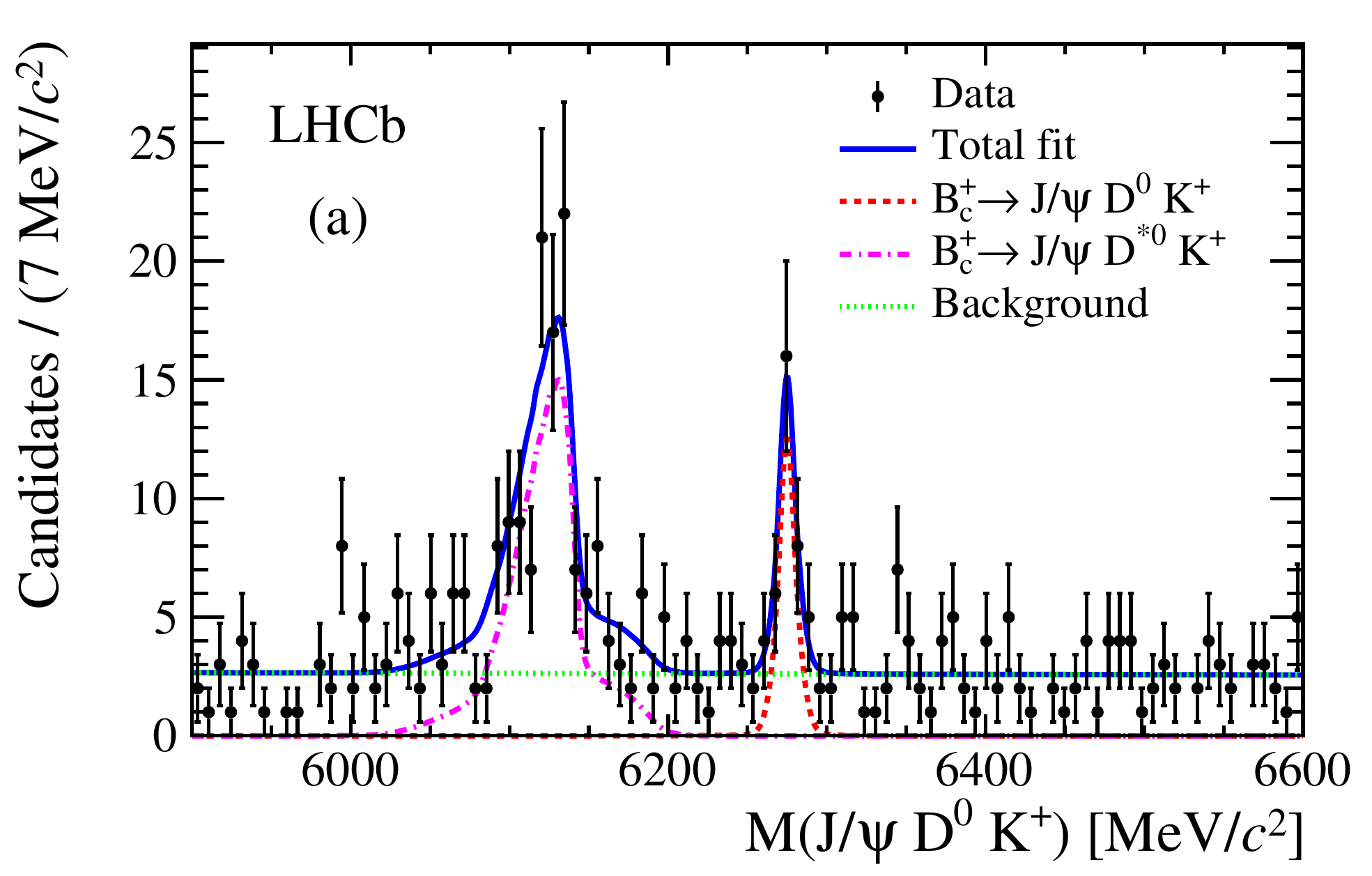}
\includegraphics[width=0.48\textwidth]{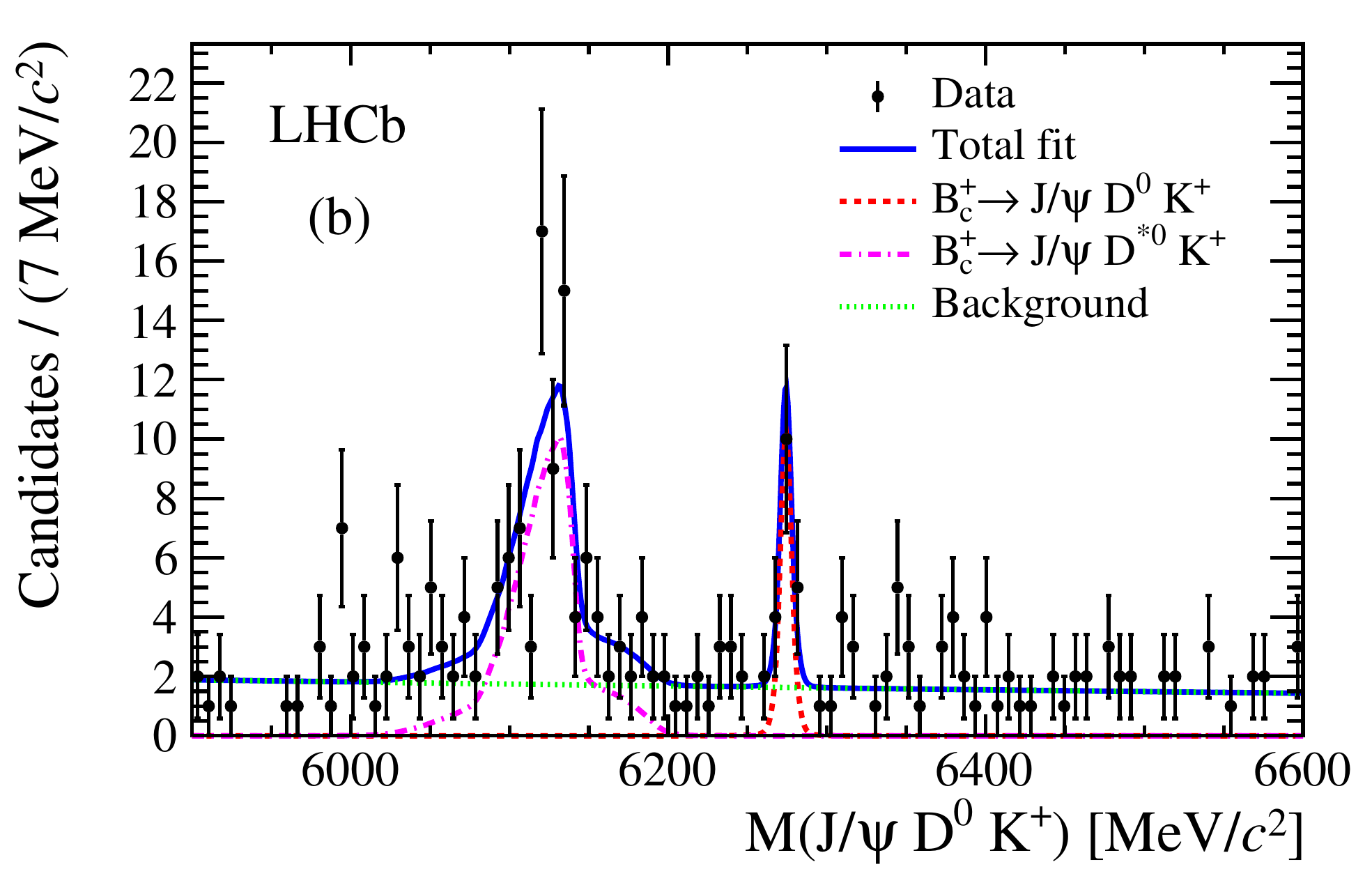}
\caption{The invariant mass distribution of $\jpsi\Dz\Kp$ candidates: 
(a) $\Dz\to\Km\pip$ and $\Dz\to\Km\pip\pim\pip$ combined; 
(b) $\Dz\to\Km\pip$ only, and the events are required to be TOS.}
\label{fig:psiDK}
\end{figure}

The invariant mass spectrum of the selected $\jpsi\Dz\Kp$ candidates is
shown in Fig.~\ref{fig:psiDK}(a), where both $\Dz\to\Km\pip$ and
$\Dz\to\Km\pip\pim\pip$ samples are combined.
The result of an extended unbinned maximum likelihood fit is also shown.
The sharp peak at the $\Bcp$ mass is the fully reconstructed $\Bcp\to\jpsi\Dz\Kp$
signal, which is fitted with the sum of a Gaussian function and 
a double-sided Crystal Ball function (DSCB),
a modified Gaussian distribution with power-law tails on both sides,
whose tail parameters are fixed from simulation. The Gaussian and the DSCB functions are constrained to have the same mean. The width of the Gaussian component is free to vary in the fit, while the ratio of the DSCB core width over the Gaussian width is fixed to the value expected from simulation.

The wider peaking structure at lower mass is due to partially reconstructed
\mbox{$\Bcp\to\jpsi\Dstarz\Kp$} signal, which is modelled using a nonparametric shape obtained from simulated
$\Dstarz\to\Dz\gamma$ and $\Dz\piz$ decays, combined according to 
their relative branching fractions~\cite{myPDG2014}.
The combinatorial background is fitted with an exponential function.
The signal yields of $\Bcp\to\jpsi\Dz\Kp$ and $\Bcp\to\jpsi\Dstarz\Kp$ decays
are $26 \pm 7$ and $102 \pm 13$, respectively.
The signal significance, $\mathcal{S}$, is estimated using 
the change in the fit likelihood from
a background-only hypothesis to a signal-plus-background hypothesis
$\mathcal{S}=\sqrt{-2\ln(\mathcal{L}_B/\mathcal{L}_{S+B})}$~\cite{Wilks:1938dza}.
Taking into account the systematic effects discussed in Sec.~\ref{sec:br}, 
the significance of the $\Bcp\to\jpsi\Dz\Kp$ signal is $6.3\sigma$ 
and the significance of the partially reconstructed 
$\Bcp\to\jpsi\Dstarz\Kp$ signal is $10.3\sigma$.
Both are observed for the first time.
An alternative method gives a compatible significance estimation. In this method
pseudoexperiments are generated using the background-only hypothesis, 
which are then fitted using the signal-plus-background hypothesis to obtain
a cumulative probability distribution $\mathcal{P}(N \geq N_S)$
as a function of the fitted signal yield $N_S$. 
Given the actual yield from data, 
the $p$-value and signal significance can be derived.
Figure~\ref{fig:psiDK}(b) shows the same mass distribution of the
$\jpsi\Dz (\to\Km\pip) \Kp$ final state for TOS triggered events.
The mass and resolution of the $\Bcp\to\jpsi\Dz\Kp$ signal distribution
are free to vary in the fit. The fitted yields 
$N(\Bcp\to\jpsi\Dz\Kp) = 14 \pm 4$ and 
$N(\Bcp\to\jpsi\Dstarz\Kp) = 69 \pm 10$, and the mass central value
$6274.20 \pm 1.40\mevcc$ are used in the branching fraction and 
mass measurements. The quoted uncertainties are statistical.

\begin{figure}[h]
\centering
\includegraphics[width=0.45\textwidth]{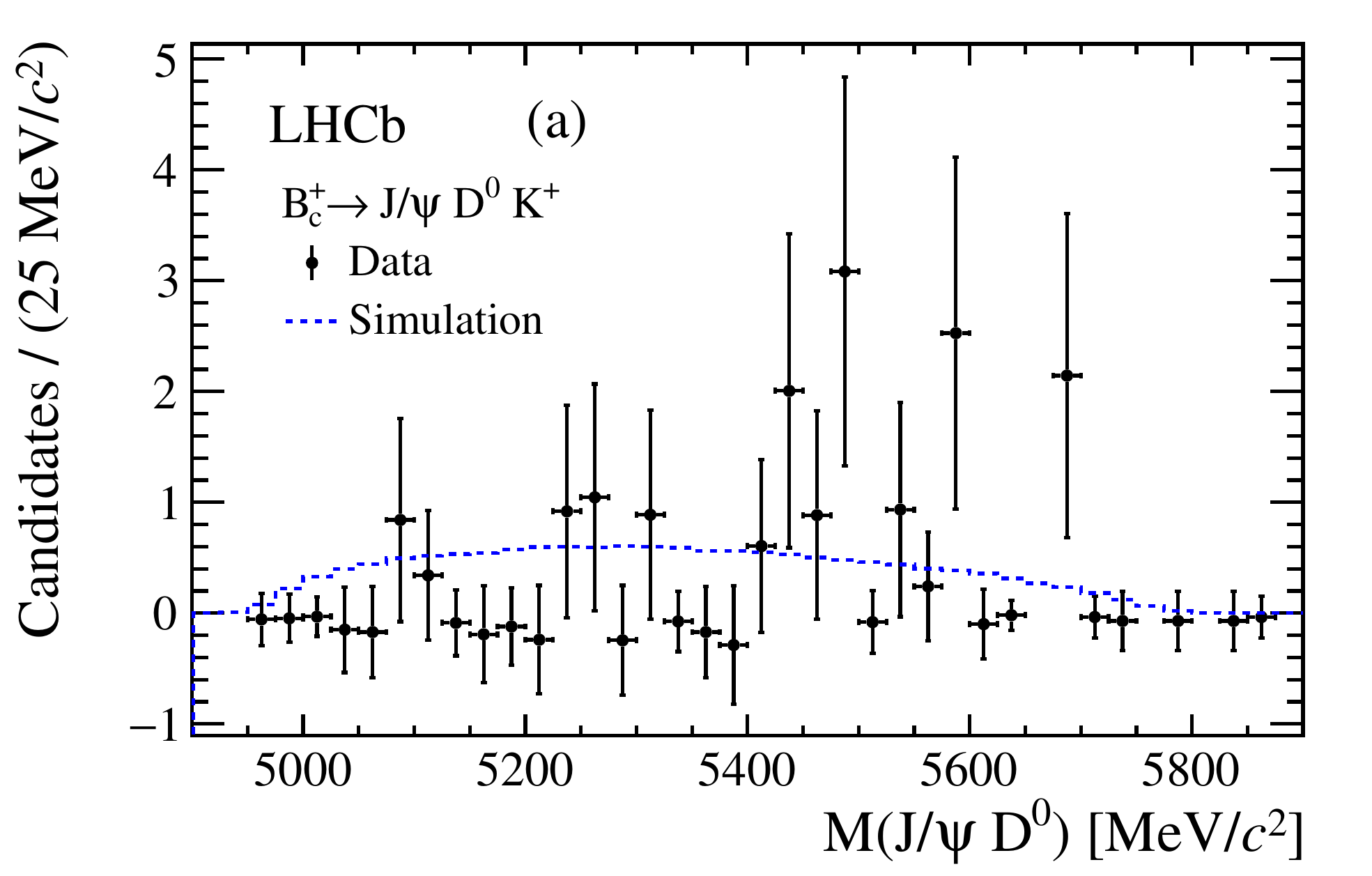}
\includegraphics[width=0.45\textwidth]{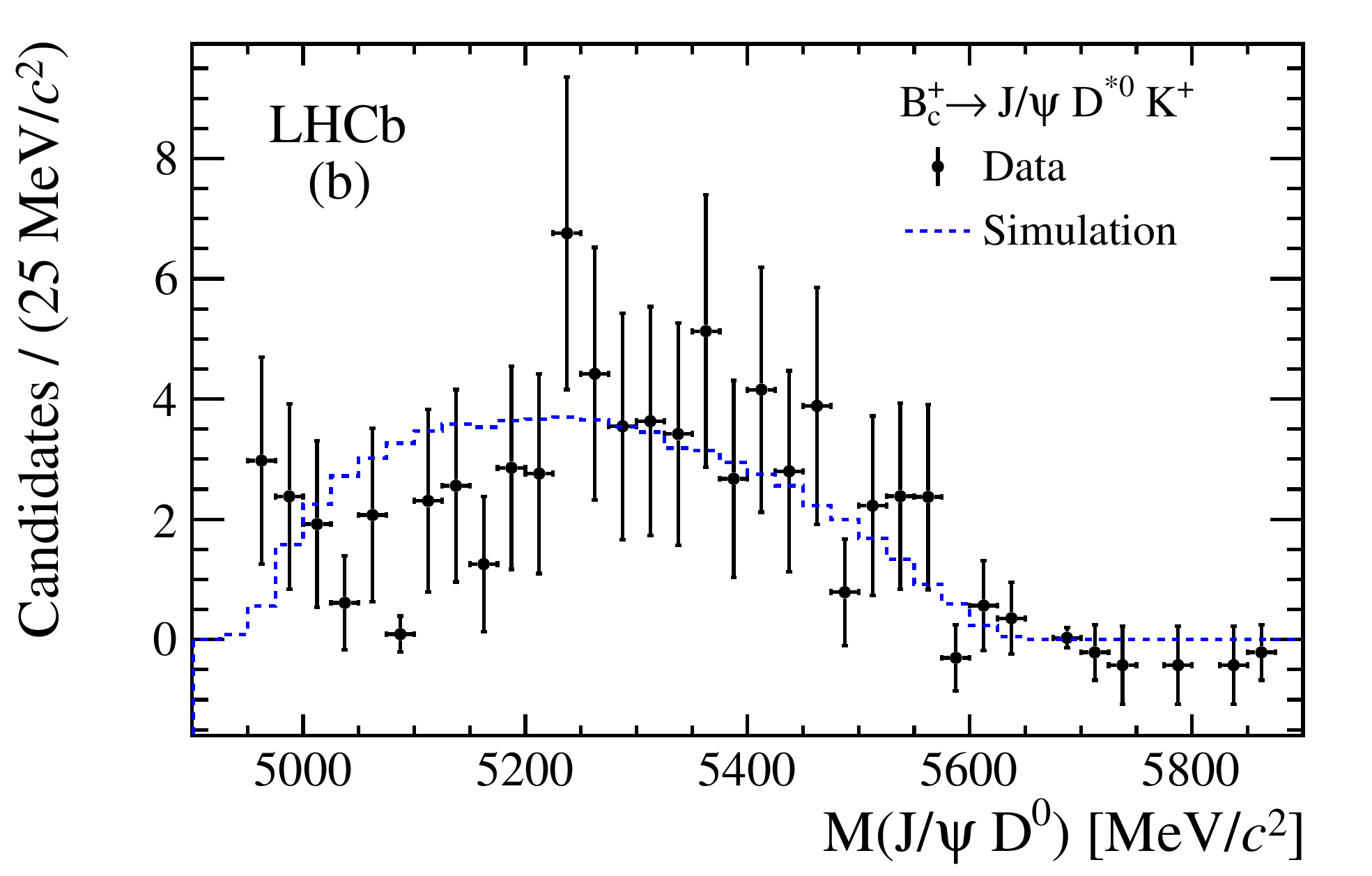}
\includegraphics[width=0.45\textwidth]{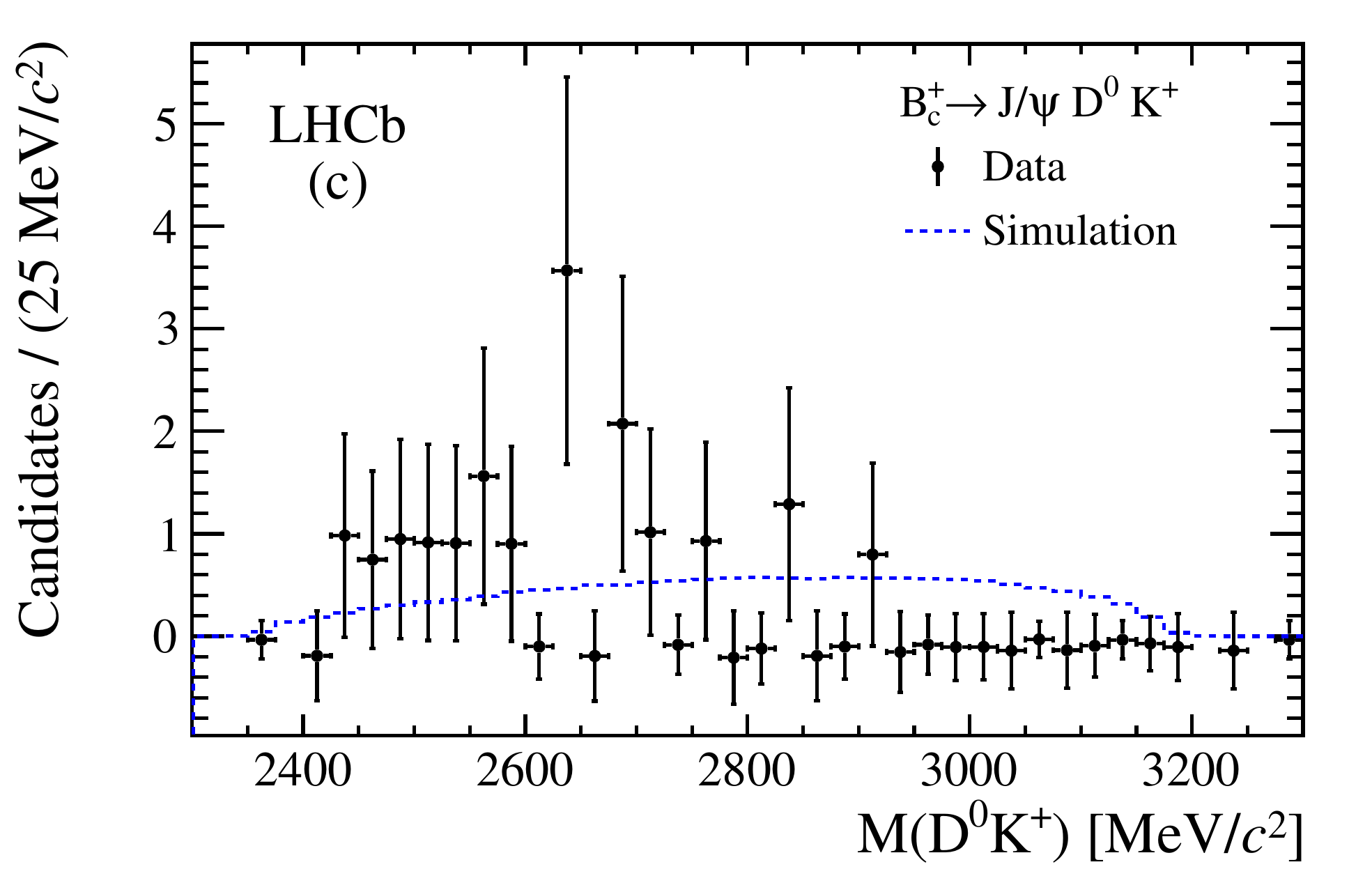}
\includegraphics[width=0.45\textwidth]{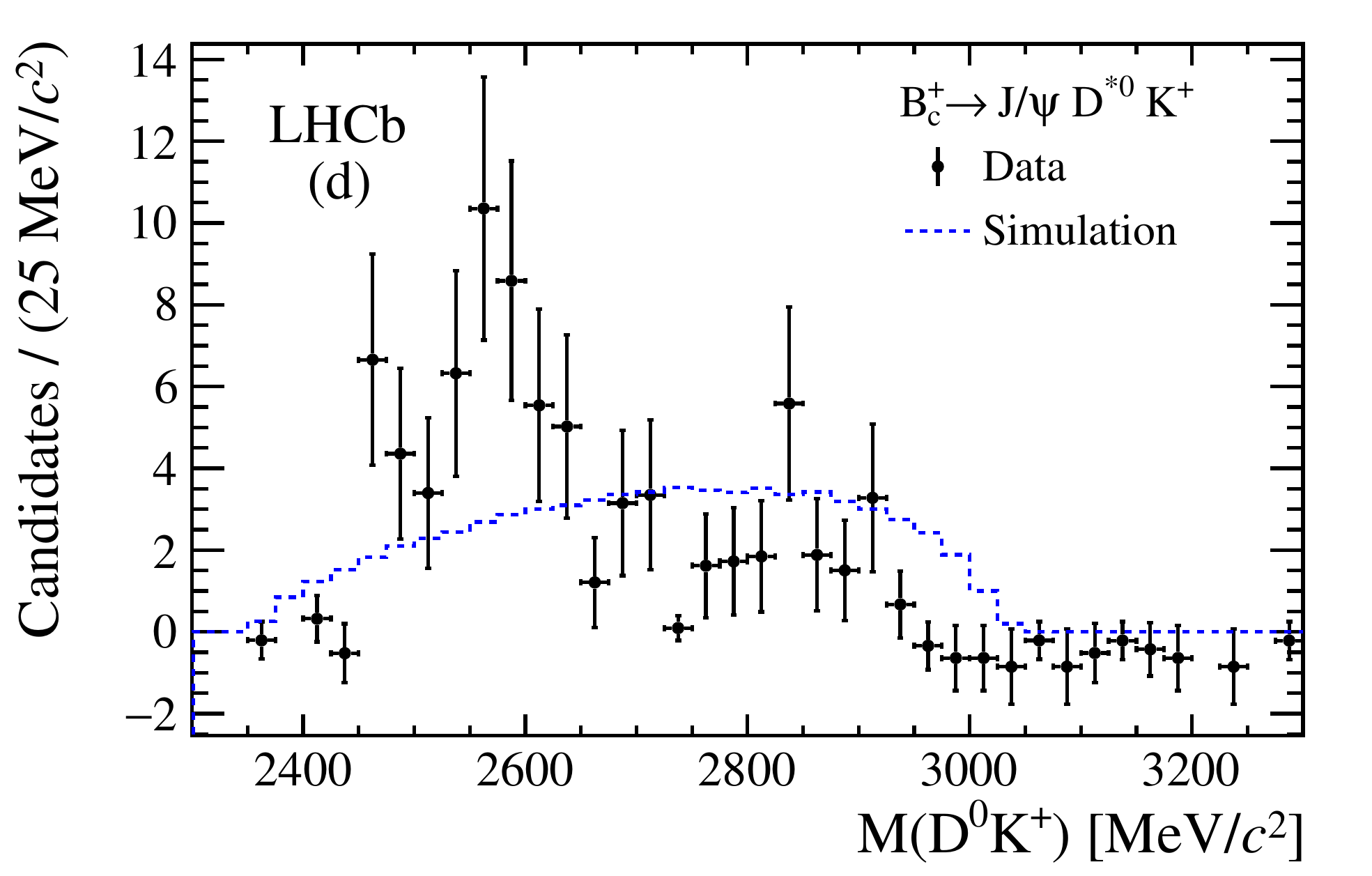}
\includegraphics[width=0.45\textwidth]{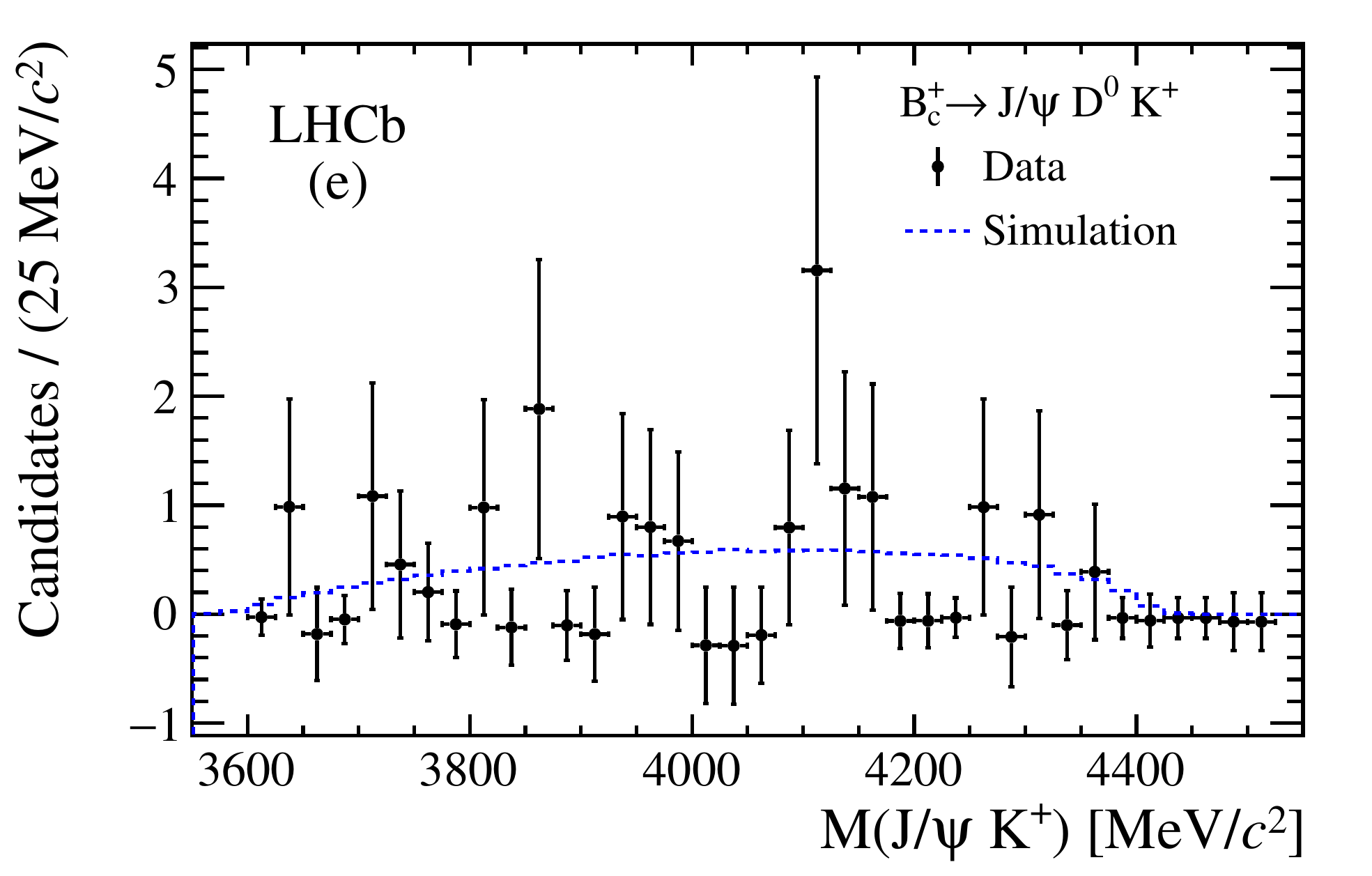}
\includegraphics[width=0.45\textwidth]{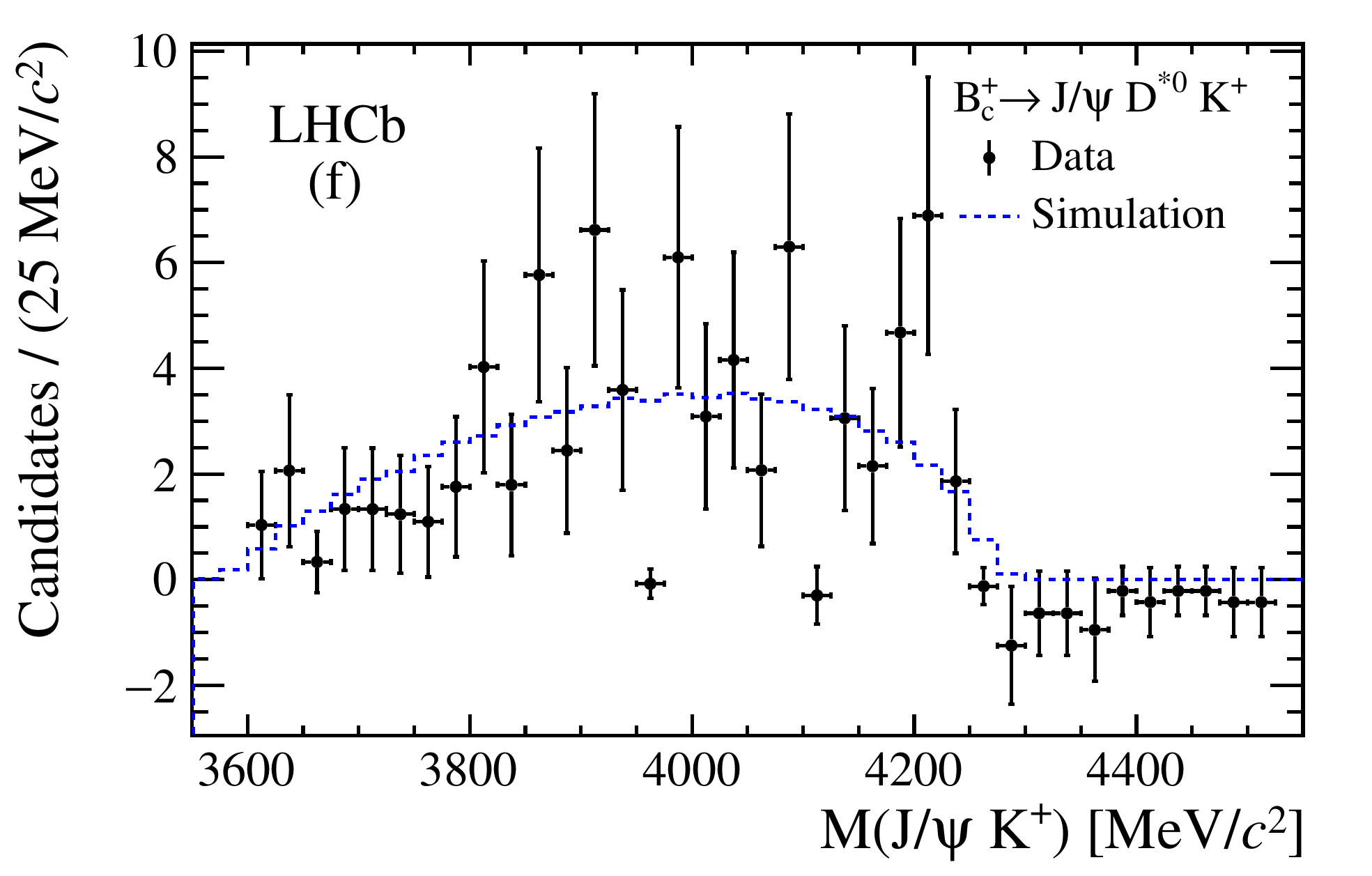}
\caption{The invariant mass distribution of (a,b) $\jpsi\Dz$, 
(c,d) $\Dz\Kp$ and (e,f) $\jpsi\Kp$ combinations of background-subtracted
(a,c,e) $\Bcp\to\jpsi\Dz\Kp$ and (b,d,f) $\Bcp\to\jpsi\Dstarz\Kp$ decays, 
where the $\gamma$ or $\piz$ in the $\Dstarz\to\Dz\gamma/\Dz\piz$ decay
is not reconstructed. Dashed lines show simulation assuming phase-space decay.}
\label{fig:decStruc}
\end{figure}

The invariant mass distributions of the $\jpsi\Dz$, $\Dz\Kp$ and $\jpsi\Kp$ 
combinations are shown in Fig.~\ref{fig:decStruc} for the $\Bcp\to\jpsi\Dz\Kp$
and $\jpsi\Dstarz\Kp$ signal events.
The background is subtracted using the \textit{sPlot} 
technique~\cite{Pivk:2004ty}, with $M(\jpsi\Dz\Kp)$ as the discriminating variable.
The distributions from simulation using a phase-space decay model
are shown for comparison.
The simulation shows comparatively poor agreement with data
for the $\Dz\Kp$ invariant mass. This distribution, sensitive to possible
intermediate resonances, should be studied further with more data.

\begin{figure}[h]
\centering
\includegraphics[width=0.48\textwidth]{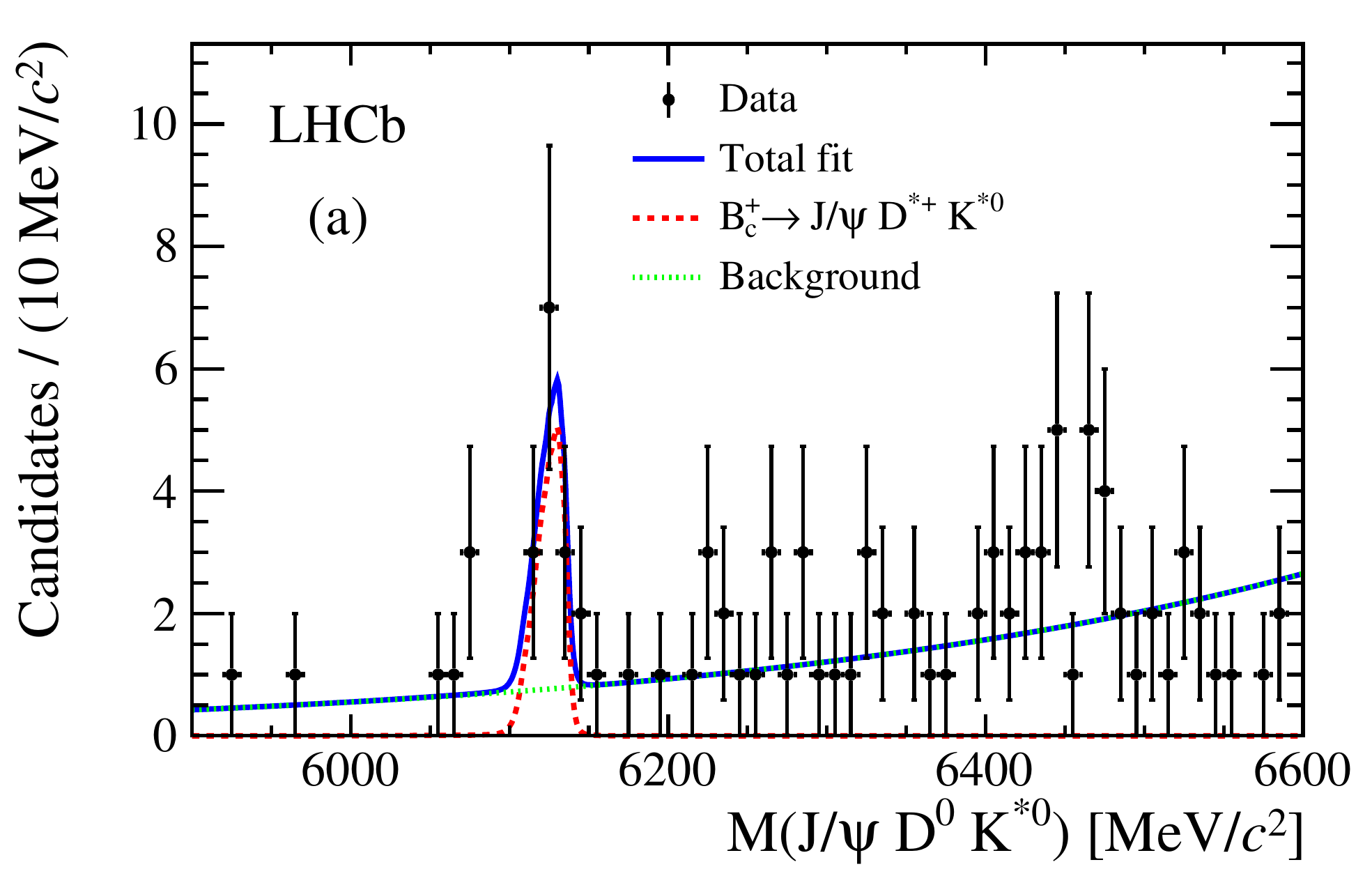}
\includegraphics[width=0.48\textwidth]{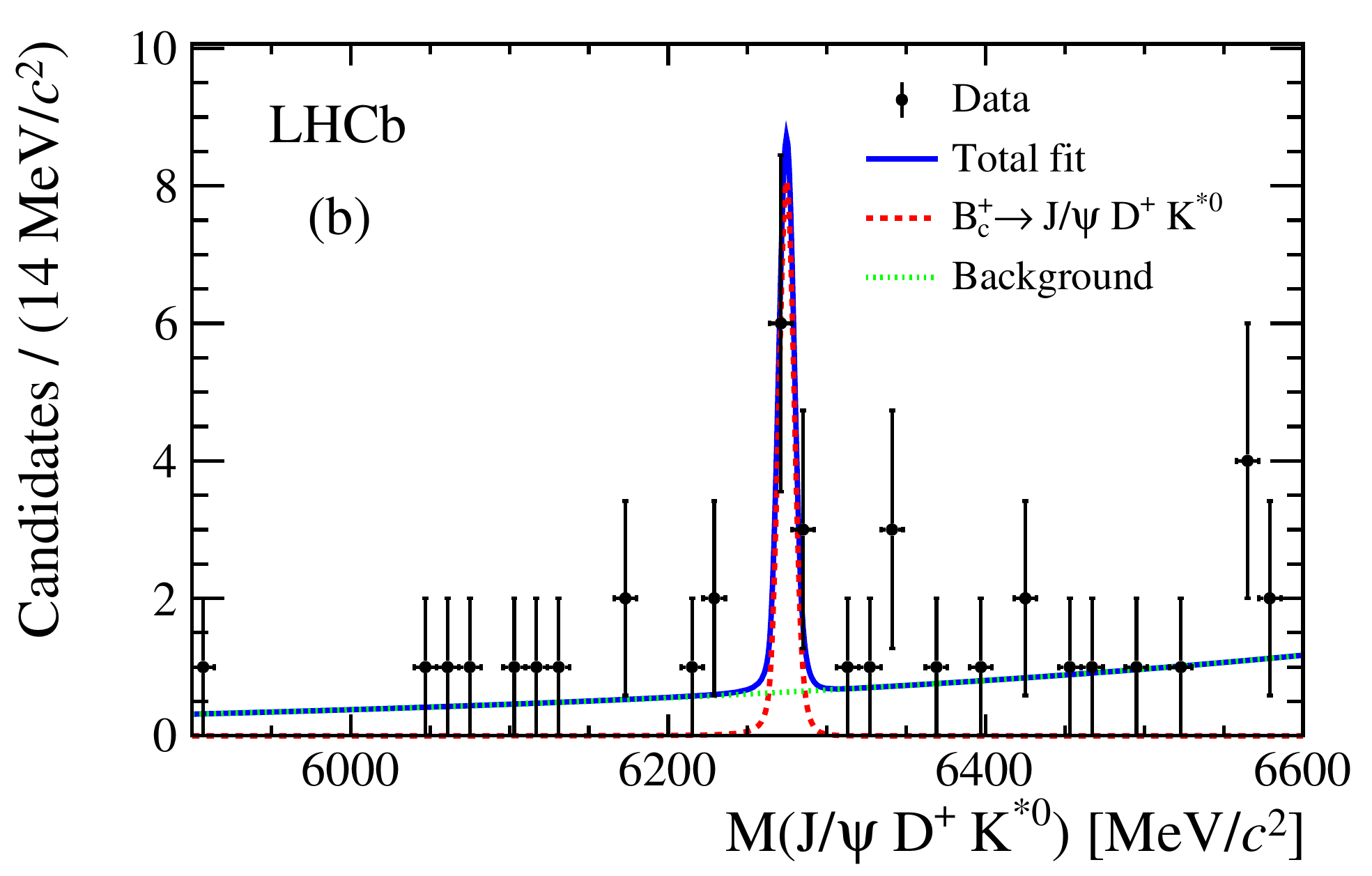}
\includegraphics[width=0.48\textwidth]{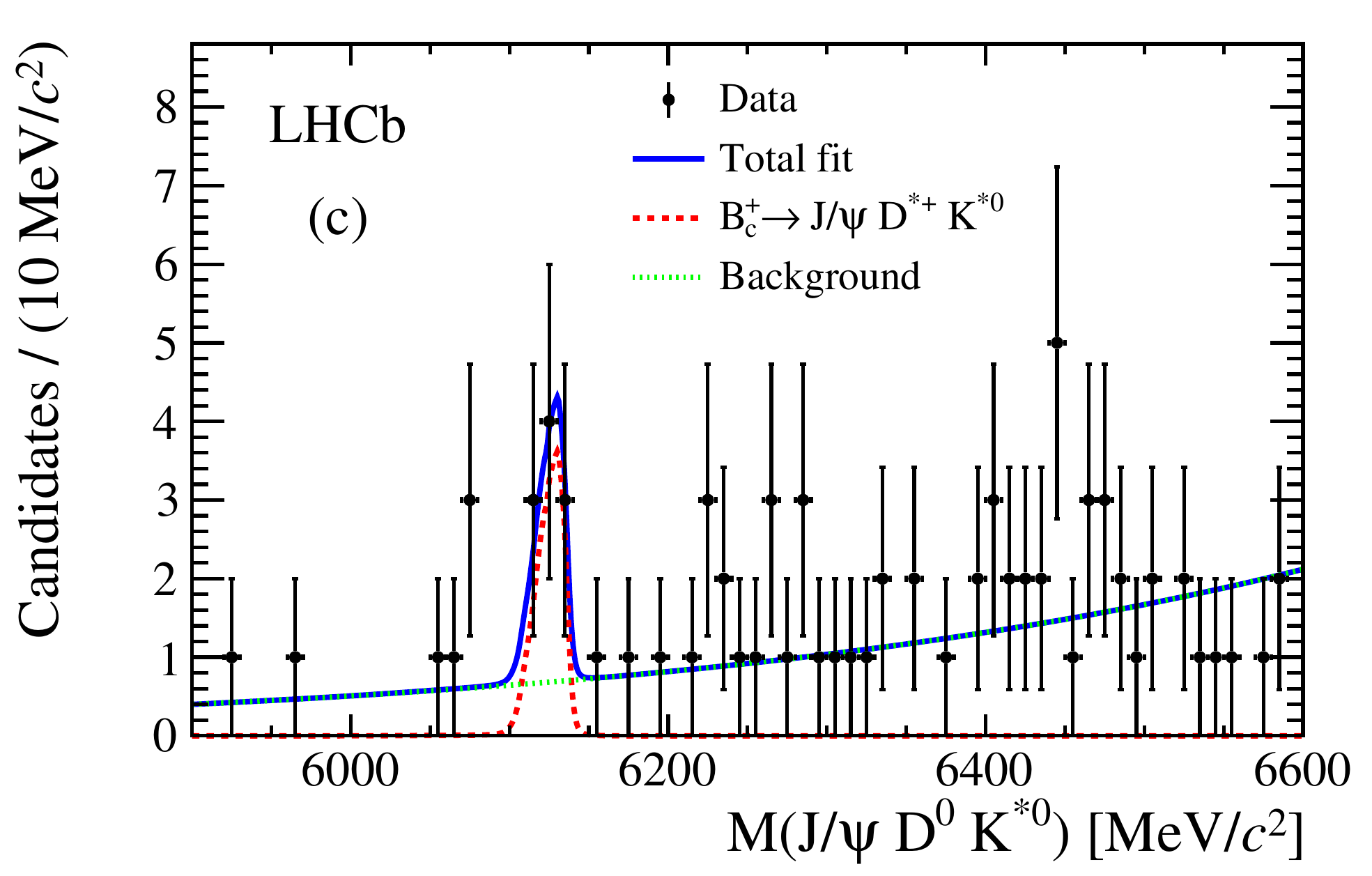}
\includegraphics[width=0.48\textwidth]{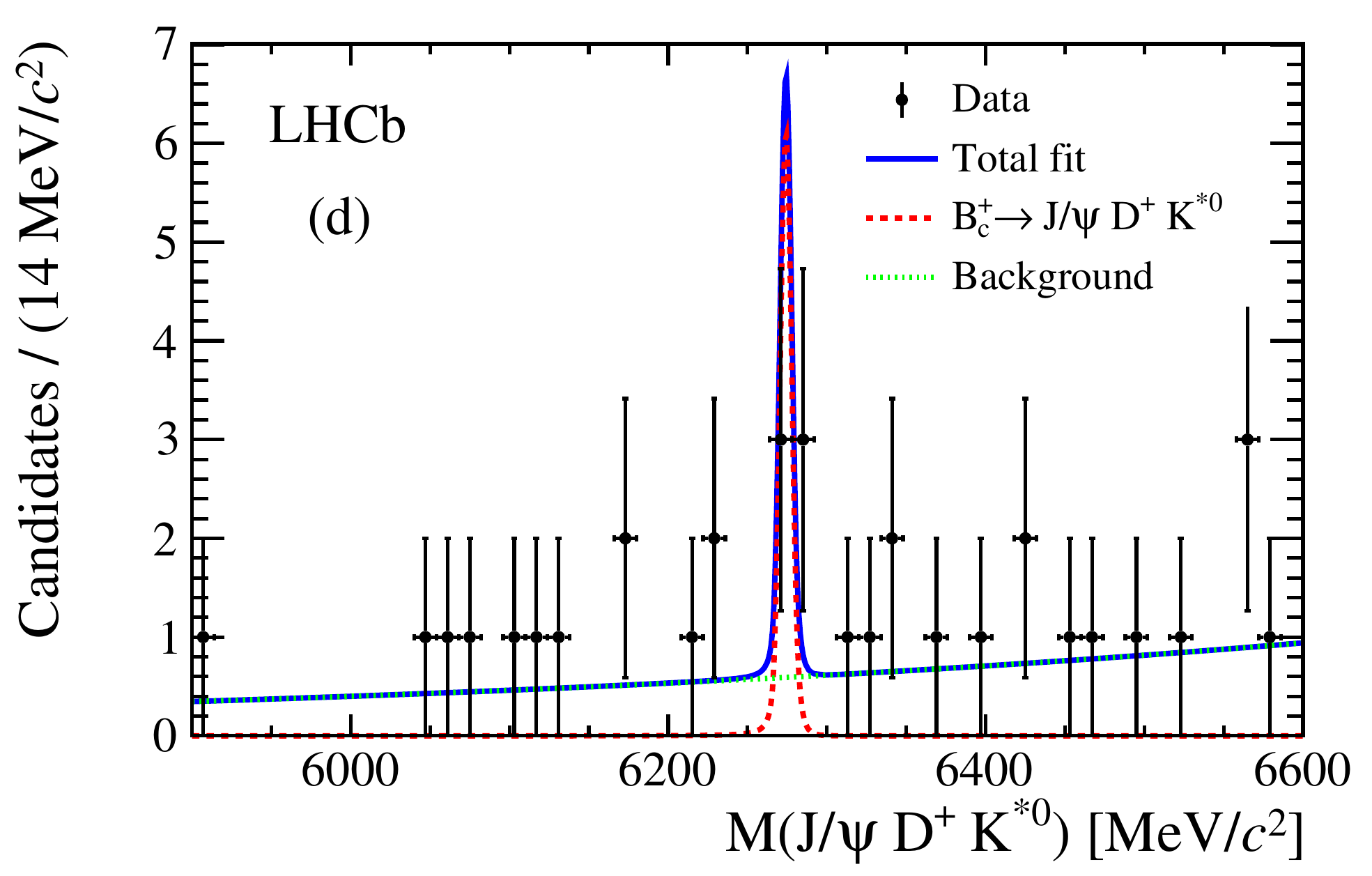}
\caption{The invariant mass distribution of (a,c) the $\Bcp\to\jpsi\Dstarp\Kstarz$ 
and (b,d) $\jpsi\Dp\Kstarz$ candidates, (a,b) without and (c,d) with TOS requirements.}
\label{fig:psiDKst-noTOS}
\end{figure}

The invariant mass distributions of the final states containing $\Kstarz$ 
candidates are shown in Fig.~\ref{fig:psiDKst-noTOS}.
The $\Bcp\to\jpsi\Dstarp\Kstarz$ decay is partially reconstructed, neglecting
the pion in the $\Dstarp\to\Dz\pip$ decay (Fig.~\ref{fig:psiDKst-noTOS}(a,c)).
The shape of the signal distribution is fixed from simulation and 
the background is modelled with an exponential function.
The $\Bcp\to\jpsi\Dp\Kstarz$ decay is fully reconstructed and modelled with 
a DSCB function, while the background is described by an exponential function 
(Fig.~\ref{fig:psiDKst-noTOS}(b,d)).
Without TOS requirements the yields of the $\Bcp\to\jpsi\Dstarp\Kstarz$ and 
$\Bcp\to\jpsi\Dp\Kstarz$ decays are $11 \pm 4$ and $7.4 \pm 2.9$ events, and
the significances are $4.0 \sigma$ and $4.4 \sigma$, respectively,
including systematic effects.
With TOS requirements applied, their yields are $7.8 \pm 3.2$ and $3.9 \pm 2.1$,
where the uncertainties are statistical only.

\begin{figure}[h]
\centering
\includegraphics[width=0.48\textwidth]{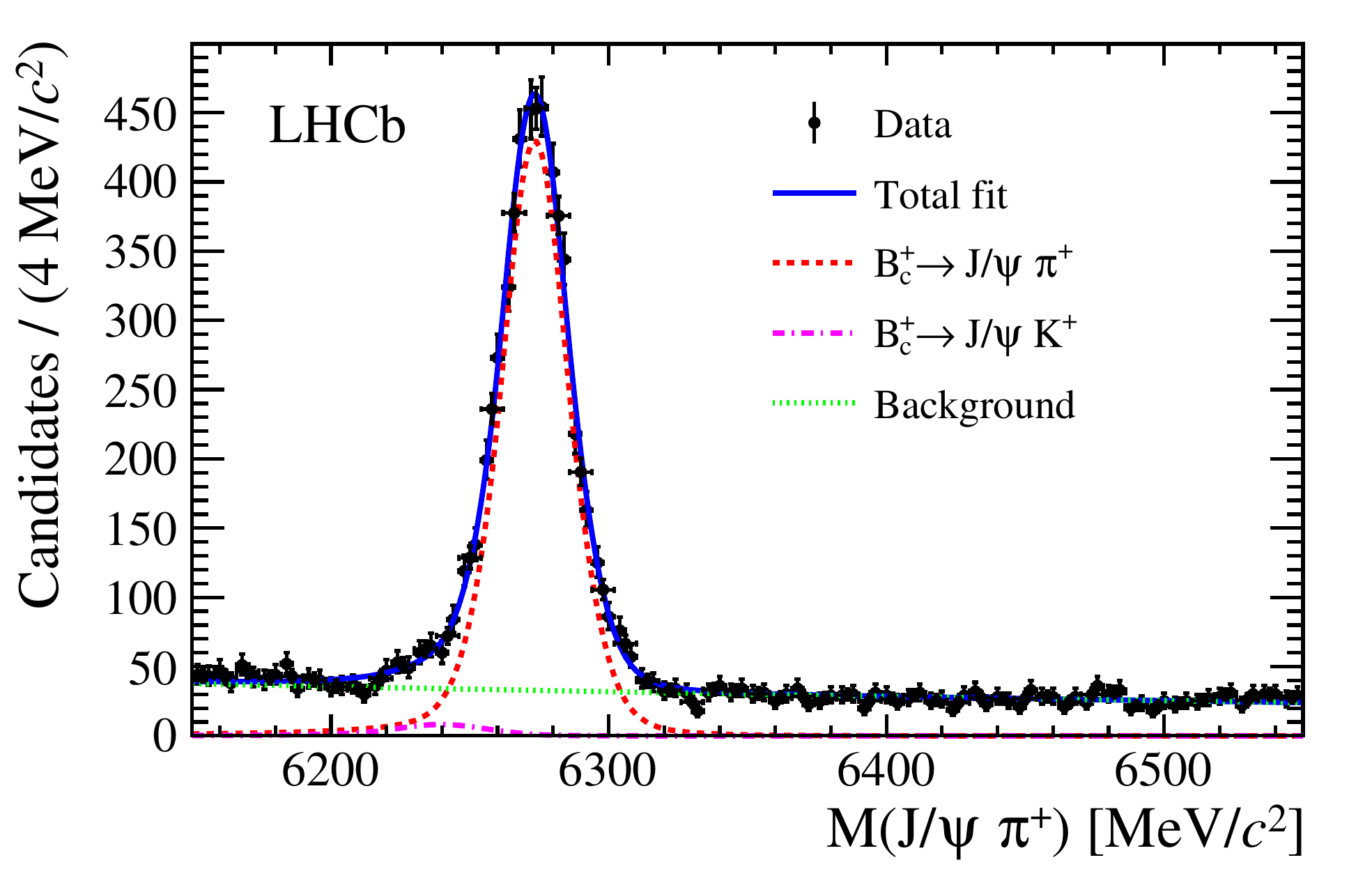}
\caption{The invariant mass distribution of the $\Bcp\to\jpsi\pip$ candidates.}
\label{fig:psipi-noTOS}
\end{figure}

The $\jpsi\pip$ mass distribution of the normalisation channel is shown
in Fig.~\ref{fig:psipi-noTOS} with TOS requirements applied.
The signal is modelled with the sum of a DSCB and a Gaussian function,
the combinatorial background with an exponential function, 
and the misidentified background from the $\Bcp\to\jpsi\Kp$ decay 
is modelled with a DSCB whose parameters are fixed
to those that describe the simulated data.
The signal yield is $3616 \pm 73$ events.

\section{Branching fraction measurement}
\label{sec:br}
After correction for detection efficiencies,
the signal yields obtained in Sec.~\ref{sec:yields} are used
to determine relative branching fractions.
The choice of the fit model is a significant source of systematic uncertainty on the signal
yield and therefore also on the branching fraction.
Alternative models are used for the signal (including a single DSCB function,
a Gaussian function, and a nonparametric shape from simulation), and the 
combinatorial background (including first- and second-order polynomial functions).
For the $\jpsi\Dz\Kp$ final state, the feed-down from higher excited 
intermediate states is considered, such as $\jpsi D_0^{*}(2400)^{0} \Kp$, 
$\jpsi \D_1(2420)^0 \Kp$ and $\chi_{c1}(\to\jpsi\gamma)\Dz\Kp$.
If these contributions, with shapes estimated by simulation, are included in the fit, 
the branching fractions change by no more than 0.5\%.
The shape of partially reconstructed $\Bcp\to\jpsi\Dstarz\Kp$ signal 
depends on the polarization and intermediate resonances in the decay.
Extreme cases of helicity amplitude configurations are generated for the decay
$\Bcp\to\jpsi \D_{\squark1}(2536)^+ (\to\Dstarz\Kp)$ and it is found that 
the unknown polarization and decay structure can change  
the signal yield by up to 5.2\%.
A dedicated simulation study shows that 
the possible peaking background from the charmless $\Bcp\to\jpsi\Kp\Km\pip$
decay~\cite{LHCb-PAPER-2013-047} is negligible.
Additionally, the fits are repeated in different mass ranges.
In the $\Bcp\to\jpsi\Dstarp\Kstarz$ sample, the background level is slightly high
around $6450 \mevcc$, but consistent with a statistical fluctuation. 
A fit in a more narrow range excluding this region gives a compatible result.
The total uncertainties due to fit modelling are found in Table~\ref{tab:br-syst} for each of the channels.

The total efficiencies are given by the product of three factors: the geometric detector 
acceptance, the reconstruction and selection efficiencies, and the trigger efficiency.
They are generally estimated using simulated samples, corrected to match the data 
when the simulation is known to be imperfect. In the simulation
the $\Bcp$ meson is generated with a lifetime of $450\fs$ taken from 
an early world average with a large uncertainty~\cite{myPDG2014}. 
For the efficiency estimation
the simulated events are therefore weighted to obtain the same lifetime 
($\tau = 511.4\fs$) as the recent and more precise LHCb
measurements~\cite{LHCb-PAPER-2013-063,LHCb-PAPER-2014-060}.
The lifetime is varied by one standard deviation ($9.3\fs$) to 
study the corresponding systematic effect, which is found to be negligible.
The simulation assumes a phase-space decay of the $\Bcp\to\jpsi\Dmesons\Kmesons$
averaged over all possible polarization configurations, and
without any intermediate decay structure.
The efficiency dependence on the invariant mass of the $D\Kmesons$ system is 
studied and the efficiencies of selected candidates are corrected event-by-event according to
the $M(D\Kmesons)$ value.
The distributions of variables used in the BDT training are compared between 
simulation and background-subtracted data, and show good agreement.
The tracking and PID efficiencies are determined in bins of track momenta,
pseudorapidity and event multiplicity using a data-driven 
method~\cite{LHCb-DP-2013-002}. 
The tracking efficiency uncertainty is estimated to be 0.4\%
per muon or hadron track, 
while for each hadron track an additional uncertainty of 1.4\% is assigned
due to the imperfect knowledge of the interaction with the detector material.
Alternative binning schemes of track momentum, pseudorapidity and event
multiplicity are applied to estimate the uncertainty on the PID efficiencies.
The systematic uncertainty on the trigger efficiency is determined to be 1.1\% 
from a comparison between data and simulation using a large $\jpsi$ 
sample~\cite{LHCb-PAPER-2013-010,LHCb-PAPER-2014-039}.
The limited data size of the simulation samples introduces systematic
uncertainties of less than 1\%. 
The uncertainties of intermediate $D^{(*)}$ decay branching 
fractions~\cite{myPDG2014} are propagated into the final results. 
Cross-checks have been performed to ensure the robustness of the results,
such as confirming that the BDT output is not correlated with the $\Bcp$ candidate mass.

The relative branching fractions of the $\Bcp$ decays are measured to be
\begin{align*}
\frac{\BF(\Bcp\to\jpsi\Dz\Kp)}{\BF(\Bcp\to\jpsi\pip)}\phantom{ii} &= 0.432 \pm 0.136 \pm 0.028, \nonumber \\
\frac{\BF(\Bcp\to\jpsi\Dstarz\Kp)}{\BF(\Bcp\to\jpsi\Dz\Kp)}\phantom{i} &= 5.1  \pm 1.8  \pm 0.4, \nonumber \\
\frac{\BF(\Bcp\to\jpsi\Dstarp\Kstarz)}{\BF(\Bcp\to\jpsi\Dz\Kp)} &= 2.10 \pm 1.08 \pm 0.34, \nonumber \\
\frac{\BF(\Bcp\to\jpsi\Dp\Kstarz)}{\BF(\Bcp\to\jpsi\Dz\Kp)}\phantom{i} &= 0.63 \pm 0.39 \pm 0.08, \nonumber 
\end{align*}
where the first uncertainty is statistical and the second is systematic.
The systematic uncertainties are summarised in Table~\ref{tab:br-syst}.
\begin{table}[t]
\small
\begin{center}
\caption{Summary of systematic uncertainties on the ratios of the $\Bcp$ decay branching fractions, in \%.}
\label{tab:br-syst}
\begin{tabular}{@{}lcccc@{}}
\hline\noalign{\smallskip}
Source of uncertainty
 & $\frac{\BF(\jpsi\Dz\Kp)}{\BF(\jpsi\pip)}$
 & $\frac{\BF(\jpsi\Dstarz\Kp)}{\BF(\jpsi\Dz\Kp)}$
 & $\frac{\BF(\jpsi\Dstarp\Kstarz)}{\BF(\jpsi\Dz\Kp)}$
 & $\frac{\BF(\jpsi\Dp\Kstarz)}{\BF(\jpsi\Dz\Kp)}$\\
\noalign{\smallskip}
\hline
Fit model  & 2.6 & 6.6 & 15.6 & 10.7 \\
Decay structure  & 1.8 & 2.2 & \phantom{1}3.1 & \phantom{1}2.9 \\
Trigger  & 1.1 & 1.1 & \phantom{1}1.1 & \phantom{1}1.1 \\
Tracking  & 2.9 & 0.0 & \phantom{1}1.5 & \phantom{1}2.9 \\
Particle identification  & 4.5 & 0.1 & \phantom{1}2.3 & \phantom{1}1.4 \\
$\Dmesons$ decay branching ratios  & 1.3 & 1.4 & \phantom{1}0.7 & \phantom{1}2.5 \\
Simulation statistics  & 0.4 & 0.6 & \phantom{1}0.8 & \phantom{1}1.0 \\
\hline
Total (\%)  & 6.5 & 7.2 & 16.2 & 11.9 \\
\hline
\end{tabular}
\end{center}
\end{table}

\section{Mass measurement}
\label{sec:mass}
The $\Bcp$ mass is determined from the fit to the $\Bcp\to\jpsi\Dz(\to\Km\pip)\Kp$
signal as shown in Fig.~\ref{fig:psiDK}(b).
The summary of systematic uncertainties is given in Table~\ref{tab:mass-syst}.
The dominant term is the momentum scale calibration. 
For a mass measurement, the momenta of the final-state particles need to be 
measured precisely.
In previous studies a large sample of $\Bp\to\jpsi\Kp, \jpsi\to\mup\mun$ decays
was used to calibrate the track momentum, and the uncertainty on 
the momentum scale calibration was determined to be 
0.03\%~\cite{LHCb-PAPER-2013-011}. 
This causes a change in the central value of the $\Bcp$ mass
by up to $0.26 \mevcc$. 
Using the same procedure as described in Sec.~\ref{sec:br},
the choice of the model is estimated to introduce an uncertainty of $0.18 \mevcc$.
\begin{table}[b]
\centering
\caption{Summary of systematic uncertainties of the \Bcp mass measurement.}
\label{tab:mass-syst}
\begin{tabular}{@{}lc@{}}
\hline
Source & Uncertainty (\mevcc) \\
\hline
Momentum scale & 0.26 \\
Fit model & 0.18 \\
Final-state radiation & 0.01 \\
$\Dz$, $\jpsi$ mass uncertainties & 0.05 \\
Energy loss correction & 0.05 \\
\hline
Total & 0.32 \\
\hline
\end{tabular}
\end{table}
The effect of soft photon emission via final-state radiation is minimised by
constraining the reconstructed $\jpsi$ and $\Dz$ masses to their nominal values.
Any remaining bias is investigated using a large sample of simulated pseudoexperiments,
which results in a correction of $+0.08 \mevcc$ to the central value,
with an uncertainty of $0.01 \mevcc$.
The uncertainties associated with the $\jpsi$ ($0.006 \mevcc$)
and $\Dz$ ($0.05 \mevcc$)
masses~\cite{PDG2016} are propagated to the $\Bcp$ mass.
The effect of an imperfect energy loss correction has been studied in the previous
$b$-hadron mass measurements~\cite{LHCb-PAPER-2011-035}
by varying the amount of detector material.
The corresponding uncertainty is $0.05 \mevcc$ for the $\Bcp$ mass measurement.
The $\Bcp$ mass is determined to be $6274.28 \pm 1.40 \pm 0.32 \mevcc$,
consistent with previous LHCb results~\cite{LHCb-PAPER-2012-028,LHCb-PAPER-2013-010,LHCb-PAPER-2014-039} and the world average~\cite{PDG2016}.
This is the most precise single measurement of the $\Bcp$ mass.
Including this result, the new LHCb average is $6274.6 \pm 1.0 \mevcc$,
where the correlated systematic uncertainties between the measurements
including those due to momentum scale and energy loss corrections
are fully accounted for.

\section{Conclusion}
\label{sec:summary}
The decays $\Bcp\to\jpsi\Dz\Kp$ and $\Bcp\to\jpsi\Dstarz\Kp$ are observed for the first time 
with $pp$ collision data corresponding to an integrated luminosity of $3 \invfb$,
collected by the LHCb experiment at centre-of-mass energies of 7 and $8 \tev$.
First evidence is reported for the $\Bcp\to\jpsi\Dstarp\Kstarz$ and
$\jpsi\Dp\Kstarz$ decays.
The $\Bc\to\jpsi\Dz\Kp$ branching fraction is measured relative to the
$\Bcp\to\jpsi\pip$ decay, and all the other signal channels are measured 
relative to the $\Bcp\to\jpsi\Dz\Kp$ decay.
The $\Bcp\to\jpsi\Dmesons\Kp$ decay has significant potential for studies of
excited \Ds states when more data are recorded.
The $\Bcp$ mass is measured to be $6274.28 \pm 1.40 \pm 0.32 \mevcc$, 
which is the most precise single measurement and 
is in good agreement with the world average and the previous LHCb results. In combination with previous results by the LHCb~\cite{LHCb-PAPER-2012-028, LHCb-PAPER-2013-010, LHCb-PAPER-2014-039} experiment, the $\Bcp$ mass is determined to be $6274.6 \pm 1.0 \mevcc$.

\section*{Acknowledgements}
 
\noindent We express our gratitude to our colleagues in the CERN
accelerator departments for the excellent performance of the LHC. We
thank the technical and administrative staff at the LHCb
institutes. We acknowledge support from CERN and from the national
agencies: CAPES, CNPq, FAPERJ and FINEP (Brazil); NSFC (China);
CNRS/IN2P3 (France); BMBF, DFG and MPG (Germany); INFN (Italy); 
FOM and NWO (The Netherlands); MNiSW and NCN (Poland); MEN/IFA (Romania); 
MinES and FASO (Russia); MinECo (Spain); SNSF and SER (Switzerland); 
NASU (Ukraine); STFC (United Kingdom); NSF (USA).
We acknowledge the computing resources that are provided by CERN, IN2P3 (France), KIT and DESY (Germany), INFN (Italy), SURF (The Netherlands), PIC (Spain), GridPP (United Kingdom), RRCKI and Yandex LLC (Russia), CSCS (Switzerland), IFIN-HH (Romania), CBPF (Brazil), PL-GRID (Poland) and OSC (USA). We are indebted to the communities behind the multiple open 
source software packages on which we depend.
Individual groups or members have received support from AvH Foundation (Germany),
EPLANET, Marie Sk\l{}odowska-Curie Actions and ERC (European Union), 
Conseil G\'{e}n\'{e}ral de Haute-Savoie, Labex ENIGMASS and OCEVU, 
R\'{e}gion Auvergne (France), RFBR and Yandex LLC (Russia), GVA, XuntaGal and GENCAT (Spain), Herchel Smith Fund, The Royal Society, Royal Commission for the Exhibition of 1851 and the Leverhulme Trust (United Kingdom).


\addcontentsline{toc}{section}{References}
\setboolean{inbibliography}{true}
\ifx\mcitethebibliography\mciteundefinedmacro
\PackageError{LHCb.bst}{mciteplus.sty has not been loaded}
{This bibstyle requires the use of the mciteplus package.}\fi
\providecommand{\href}[2]{#2}

\newpage

\centerline{\large\bf LHCb collaboration}
\begin{flushleft}
\small
R.~Aaij$^{40}$,
B.~Adeva$^{39}$,
M.~Adinolfi$^{48}$,
Z.~Ajaltouni$^{5}$,
S.~Akar$^{59}$,
J.~Albrecht$^{10}$,
F.~Alessio$^{40}$,
M.~Alexander$^{53}$,
S.~Ali$^{43}$,
G.~Alkhazov$^{31}$
% ...

{\footnotesize \it
% ...
$ ^{5}$Clermont Universit{\'e}, Universit{\'e} Blaise Pascal, CNRS/IN2P3, LPC, Clermont-Ferrand, France\\
% ...
$ ^{10}$Fakult{\"a}t Physik, Technische Universit{\"a}t Dortmund, Dortmund, Germany\\
% ...
$ ^{31}$Petersburg Nuclear Physics Institute (PNPI), Gatchina, Russia\\
% ...
$ ^{39}$Universidad de Santiago de Compostela, Santiago de Compostela, Spain\\
$ ^{40}$European Organization for Nuclear Research (CERN), Geneva, Switzerland\\
% ...
$ ^{43}$Nikhef National Institute for Subatomic Physics, Amsterdam, The Netherlands\\
% ...
$ ^{48}$H.H. Wills Physics Laboratory, University of Bristol, Bristol, United Kingdom\\
% ...
$ ^{53}$School of Physics and Astronomy, University of Glasgow, Glasgow, United Kingdom\\
% ...
$ ^{59}$University of Cincinnati, Cincinnati, OH, United States\\
% ...
}
\end{flushleft}
\end{document}